\documentclass{article}
\usepackage{natbib}
\usepackage{mathptmx} 
\usepackage{helvet}       
\usepackage{courier}    
\usepackage{type1cm}  
\usepackage{amsfonts}
\usepackage{amssymb}
\usepackage{xspace}
\usepackage{makeidx}         
\usepackage{multicol}        
\usepackage[bottom]{footmisc}
\usepackage{graphicx}        

\begin{document}

%
%
\let\jnl=\rmfamily
\def\refe@jnl#1{{\jnl#1}}%

\newcommand\aj{\refe@jnl{AJ}}%
\newcommand\actaa{\refe@jnl{Acta Astron.}}%
\newcommand\araa{\refe@jnl{ARA\&A}}%
\newcommand\apj{\refe@jnl{ApJ}}%
\newcommand\apjl{\refe@jnl{ApJ}}%
\newcommand\apjs{\refe@jnl{ApJS}}%
\newcommand\ao{\refe@jnl{Appl.~Opt.}}%
\newcommand\apss{\refe@jnl{Ap\&SS}}%
\newcommand\aap{\refe@jnl{A\&A}}%
\newcommand\aapr{\refe@jnl{A\&A~Rev.}}%
\newcommand\aaps{\refe@jnl{A\&AS}}%
\newcommand\azh{\refe@jnl{AZh}}%
 \newcommand\fcp{\refe@jnl{Fundam. of Cosm. Phys.}}
\newcommand\gca{\refe@jnl{GeoCh.Act}}%
\newcommand\grl{\refe@jnl{Geo.Res.Lett.}}%
\newcommand\jgr{\refe@jnl{J.Geoph.Res.}}%
\newcommand\memras{\refe@jnl{MmRAS}}%
\newcommand\jrasc{\refe@jnl{J.RoySocCan}}%
\newcommand\memsai{\refe@jnl{Mem.Soc.Astr.It.}}%
\newcommand\mnras{\refe@jnl{MNRAS}}%
\newcommand\na{\refe@jnl{New A}}%
\newcommand\nar{\refe@jnl{New A Rev.}}%
\newcommand\pra{\refe@jnl{Phys.~Rev.~A}}%
\newcommand\prb{\refe@jnl{Phys.~Rev.~B}}%
\newcommand\prc{\refe@jnl{Phys.~Rev.~C}}%
\newcommand\prd{\refe@jnl{Phys.~Rev.~D}}%
\newcommand\pre{\refe@jnl{Phys.~Rev.~E}}%
\newcommand\prl{\refe@jnl{Phys.~Rev.~Lett.}}%
\newcommand\pasa{\refe@jnl{PASA}}%
\newcommand\pasp{\refe@jnl{PASP}}%
\newcommand\pasj{\refe@jnl{PASJ}}%
\newcommand\physscr{\refe@jnl{Phys.Scr.}}%
\newcommand\planss{\refe@jnl{Planetary and Space Science}}
\newcommand\skytel{\refe@jnl{S\&T}}%
\newcommand\solphys{\refe@jnl{Sol.~Phys.}}%
\newcommand\sovast{\refe@jnl{Soviet~Ast.}}%
\newcommand\ssr{\refe@jnl{Space~Sci.~Rev.}}%
\newcommand\nat{\refe@jnl{Nature}}%
\newcommand\iaucirc{\refe@jnl{IAU~Circ.}}%
\newcommand\aplett{\refe@jnl{Astrophys.~Lett.}}%
\newcommand\apspr{\refe@jnl{Astrophys.~Space~Phys.~Res.}}%
\newcommand\nphysa{\refe@jnl{Nucl.~Phys.~A}}%
\newcommand\physrep{\refe@jnl{Phys.~Rep.}}%
\newcommand\procspie{\refe@jnl{Proc.~SPIE}}%

\newcommand{\Al}{$^{26}$Al\xspace}
\newcommand{\al}{$^{26}$Al\xspace}
\newcommand{\Be}{$^{7}$Be\xspace}
\newcommand{\be}{$^{7}$Be\xspace}
\newcommand{\bem}{$^{10}$Be\xspace}
\newcommand{\ca}{$^{44}$Ca\xspace}
\newcommand{\Ca}{$^{44}$Ca\xspace}
\newcommand{\cam}{$^{41}$Ca\xspace}
\newcommand{\Co}{$^{56}$Co\xspace}
\newcommand{\co}{$^{56}$Co\xspace}
\newcommand{\csm}{$^{135}$Cs\xspace}
\newcommand{\ct}{$^{13}$C\xspace}
\newcommand{\ci}{$^{57}$Co\xspace}
\newcommand{\Ci}{$^{57}$Co\xspace}
\newcommand{\ch}{$^{60}$Co\xspace}
\newcommand{\Ch}{$^{60}$Co\xspace}
\newcommand{\Cl}{$^{36}$Cl\xspace}
\newcommand{\li}{$^{7}$Li\xspace}
\newcommand{\Li}{$^{7}$Li\xspace}
\newcommand{\Fe}{$^{60}$Fe\xspace}
\newcommand{\fh}{$^{60}$Fe\xspace}
\newcommand{\fe}{$^{56}$Fe\xspace}
\newcommand{\Fr}{$^{57}$Fe\xspace}
\newcommand{\fr}{$^{57}$Fe\xspace}
\newcommand{\mg}{$^{26}$Mg\xspace}
\newcommand{\Mg}{$^{26}$Mg\xspace}
\newcommand{\mn}{$^{54}$Mn\xspace}
\newcommand{\Na}{$^{22}$Na\xspace}
\newcommand{\Ne}{$^{22}$Ne\xspace}
\newcommand{\Ni}{$^{56}$Ni\xspace}
\newcommand{\nh}{$^{60}$Ni\xspace}
\newcommand{\Nh}{$^{60}$Ni\xspace}
\newcommand\nuk[2]{$\rm ^{\rm #2} #1$}  
\newcommand{\pd}{$^{107}$Pd\xspace}
\newcommand{\pb}{$^{205}$Pb\xspace}
\newcommand{\tc}{$^{99}$Tc\xspace}
\newcommand{\Sc}{$^{44}$Sc\xspace}
\newcommand{\Ti}{$^{44}$Ti\xspace}
\newcommand{\ti}{$^{44}$Ti\xspace}
\def\aa{$\alpha$}
\newcommand{\about}{$\simeq$}
\newcommand{\cms}{cm\ensuremath{^{-2}} s\ensuremath{^{-1}}\xspace}
\newcommand{\degree}{$^{\circ}$}
\newcommand{\flux}{ph~cm\ensuremath{^{-2}} s\ensuremath{^{-1}}\xspace}
\newcommand{\fluxrad}{ph~cm$^{-2}$s$^{-1}$rad$^{-1}$\ }
\newcommand{\ga}{\ensuremath{\gamma}}
\newcommand{\gam}{\ensuremath{\gamma}}
\def\nn{$\nu$}
\def\ra{$\rightarrow$}
\newcommand{\Msol}{M\ensuremath{_\odot}\xspace}
\newcommand{\msol}{M\ensuremath{_\odot}\xspace}
\newcommand{\Msolppc}{M\ensuremath{_\odot} pc\ensuremath{^{-2}}{\xspace}}
\newcommand{\Msolpy}{M\ensuremath{_\odot} y\ensuremath{^{-1}}{\xspace}}
\newcommand{\mspc}{M\ensuremath{_\odot} pc\ensuremath{^{-2}}{\xspace}}
\newcommand{\msb}{M\ensuremath{_\odot}\xspace}
\newcommand{\Msun}{M\ensuremath{_\odot}\xspace}
\newcommand{\Rsun}{R\ensuremath{_\odot}\xspace}
\newcommand{\rsun}{R\ensuremath{_\odot}\xspace}
\newcommand{\Lsun}{L\ensuremath{_\odot}\xspace}
\newcommand{\lsun}{L\ensuremath{_\odot}\xspace}
\newcommand{\solar}{\ensuremath{_\odot}\xspace}
\newcommand{\zs}{Z\ensuremath{_\odot}\xspace}
\newcommand{\zsun}{Z\ensuremath{_\odot}\xspace}

\newcommand{\hmol}{H$_{\rm 2}$}
\newcommand{\hatm}{H${\rm I}$}

\newcommand{\oq}{\textquotedblleft}
\newcommand{\cq}{\textquotedblright}
\newcommand{\cqb}{{\textquotedblright}~}    
\newcommand{\hilight} [1] {{\it #1}}  


\author{Roland Diehl\footnote{Max Planck Institut f\"ur extraterrestrische Physik, 85748 Garching, Germany}
}
\title{Astrophysics with Radioactive Isotopes}
\label{intro}
\maketitle

\abstract{Radioactivity was discovered as a by-product of searching for elements with suitable chemical properties. Understanding its characteristics led to the development of nuclear physics, understanding that unstable configurations of nucleons transform into stable end products through radioactive decay. In the universe, nuclear reactions create new nuclei under the energetic circumstances characterising cosmic nucleosynthesis sites, such as the cores of stars and supernova explosions. Observing the radioactive decays of unstable nuclei, which are by-products of such cosmic nucleosynthesis, is a special discipline of astronomy. Understanding these special cosmic sites, their environments, their dynamics, and their physical processes, is the \emph{Astrophysics with Radioactivities} that makes the subject of this book. We address the history, the candidate sites of nucleosynthesis, the different observational opportunities, and the tools of this field of astrophysics.}

\section{Origin of Radioactivity}
\begin{figure} 
  \includegraphics[width=\textwidth]{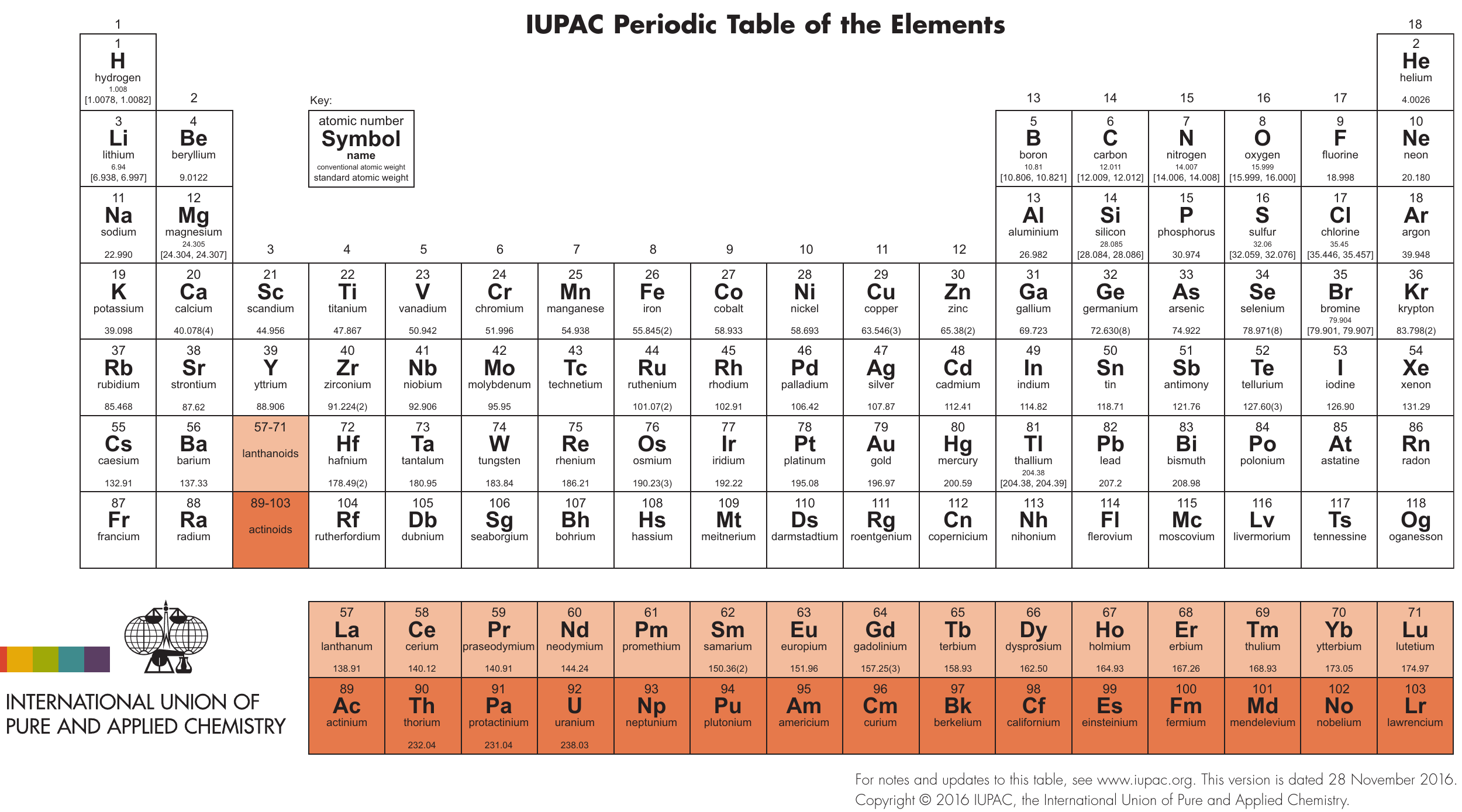}\\
  \caption{The periodic table of elements, grouping chemical elements according to their chemical-reaction properties and their atomic weight, after Mendeleyev (1869), in its 2016 version (IUPAC.org)}\label{fig_1_periodic_table}
\end{figure}

The nineteenth century spawned various efforts to bring order into the elements encountered in nature. Among the most important was an inventory of the {\it elements}  \index{elements!chemical} assembled by the Russian chemist Dimitri Mendeleyev \index{Mendeleyev, D.} in 1869, which grouped elements according to their chemical properties, their {\it valences}, as derived from the compounds they were able to form, at the same time sorting the elements by atomic weight. The genius of Mendeleyev lay in his confidence in these sorting principles, which enforce gaps in his table for expected but then unknown elements, and Mendeleyev was able to predict the physical and chemical properties of such elements-to-be-found. The tabular arrangement invented by Mendeleyev (Fig.~\ref{fig_1_periodic_table}) still is in use today, and is being populated at the high-mass end by the great experiments in heavy-ion collider laboratories to create the short-lived elements predicted to exist. The second half of the nineteenth century thus saw scientists being all-excited about chemistry and the fascinating discoveries one could make using Mendeleyev's sorting principles. Note that this was some 30~years before sub-atomic particles and the atom were discovered. Today the existence of 118 elements is firmly established\footnote{IUPAC, the international union of chemistry, coordinates definitions, groupings, and naming; see www.IUPAC.org}, the latest additions no. 113-118 all discovered in year 2016, which reflects the concerted experimental efforts.

In the late nineteenth century, scientists also were excited about new types of penetrating radiation. Conrad R\"ontgen's \index{R\"ontgen, C.} discovery in 1895 of {\it X-rays} as a type of electromagnetic radiation is important for understanding the conditions under which Antoine Henri Becquerel \index{Becquerel, A.H.} discovered radioactivity in 1896. Becquerel also was engaged in chemical experiments, in his research on phosphorescence exploiting the chemistry of photographic-plate materials. At the time, Becquerel had prepared some plates treated with uranium-carrying minerals, but did not get around to make the planned experiment. When he found the plates in their dark storage some time later, he accidentally processed them, and was surprised to find an image of a coin which happened to have been stored with the plates. Excited about X-rays, he believed he had found yet another type of radiation. Within a few years, Becquerel with Marie and Pierre Curie \index{Curie, M.} and others recognised that the origin of the observed radiation were elemental transformations of the uranium minerals: The physical process of {\it radioactivity} had been found! The revolutionary aspect of elements being able to spontaneously change their nature became masked at the beginning of the twentieth century, when sub-atomic particles and the atom were discovered. But well before atomic and quantum physics began to unfold, the physics of {\it weak interactions} had already been discovered in its form of {\it radioactivity}.

The different characteristics of different chemical elements and the systematics of Mendeleyev's periodic table were soon understood from the atomic structure of a compact and positively charged nucleus and a number of electrons orbiting the nucleus and neutralising the charge of the atom. Bohr's atomic model led to the dramatic developments of quantum mechanics and spectroscopy of atomic shell transitions. But already in 1920, Ernest Rutherford proposed that an electrically neutral particle of similar mass as the hydrogen nucleus (proton) was to be part of the compact atomic nucleus. It took more than two decades to verify by experiment the existence of this 'neutron', by James Chadwick in 1932. The atomic nucleus, too, was seen as a quantum mechanical system composed of a multitude of particles bound by the strong nuclear force. This latter characteristic is common to 'hadrons', i.e. the electrically charged proton and the neutron, the latter being slightly more massive\footnote{The mass difference is \citep{Patrignani:2016} 1.293332 MeV = 939.565413 - 938.272081 MeV for the mass of neutron and proton, respectively. One may think of the proton as the lowest-energy configuration of a hadron, that is the target of matter in a higher state, such as the combined proton-electron particle, more massive than the proton by the electron mass plus some binding energy of the quark constituents of hadrons.}. Neutrons remained a mystery for so long, as they are unstable and decay with a mean life of 880 seconds from the weak interaction into a proton, an electron, and an anti-neutrino. This is the origin of radioactivity.

The chemical and physical characteristics of an element are dominated by their electron configuration, hence by the number of charges contained in the atomic electron cloud, which again is dictated by the charge of the atomic nucleus, the number of protons. The number of neutrons included in the nucleus are important as they change the mass of the atom, however the electron configuration and hence the properties are hardly affected. Therefore, we distinguish \emph{isotopes} of each particular chemical element, which are different in the number of neutrons included in the nucleus, but carry the same charge of the nucleus. For example, we know of three stable isotopes of oxygen as found in nature, $^{16}$O,  $^{17}$O, and $^{18}$O. There are more possible nucleus configurations of oxygen with its eight protons, ranging from $^{13}$O as the lightest and $^{24}$O as the most massive known isotope.

\begin{figure} 
  \includegraphics[width=\textwidth]{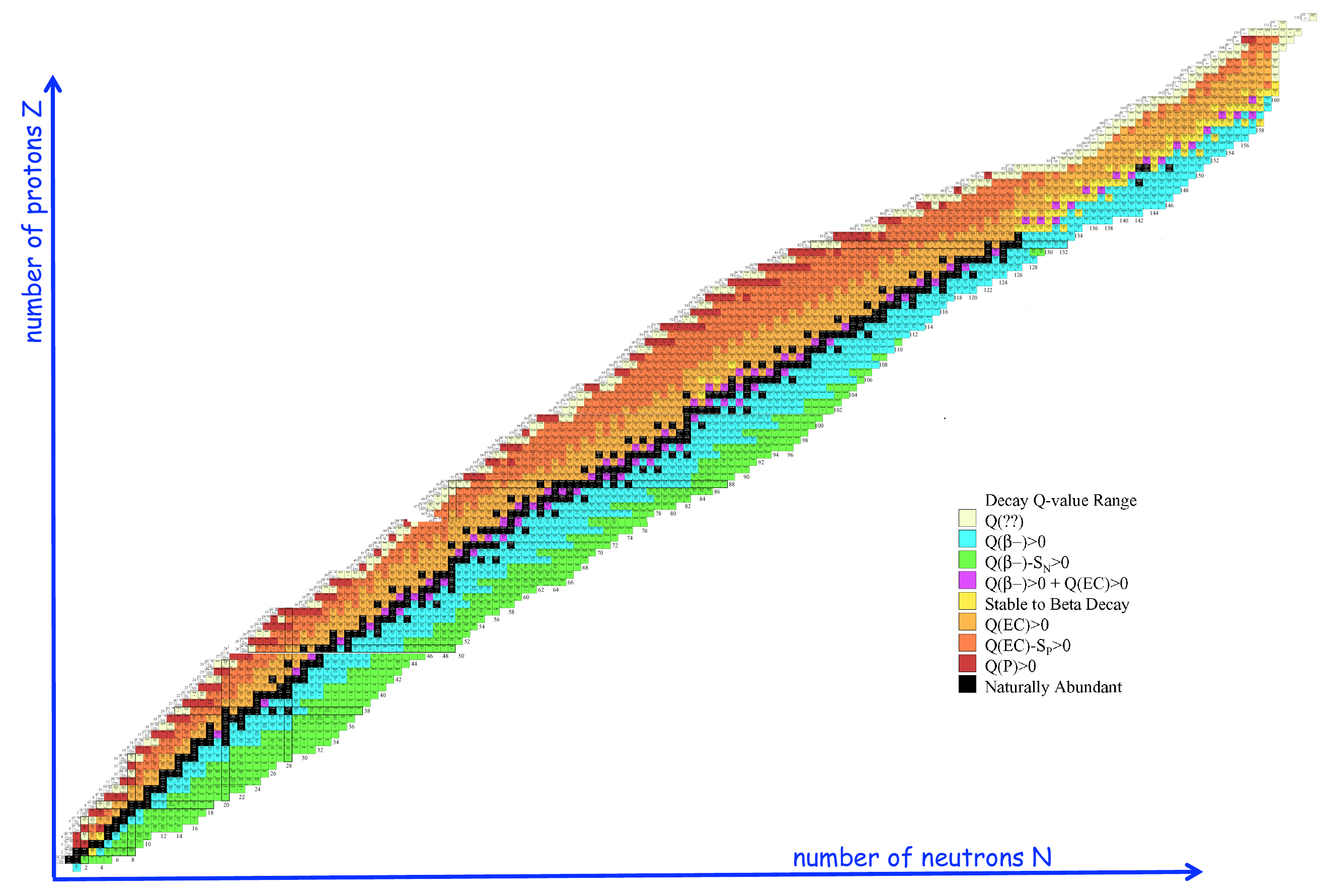}\\
  \caption{The table of isotopes, showing nuclei in a chart of neutron number (abscissa) versus proton number (ordinate). The stable elements are marked in black.
  All other isotopes are unstable, or radioactive, and will decay until a stable nucleus is obtained.}
  \label{fig_1_table-of-isotopes}
\end{figure}

An {\it isotope} is defined by the number of its  two types of nucleons\footnote{The sub-atomic particles in the nucleus are composed of three quarks, and also called \emph{baryons}. Together with the two-quark particles called \emph{mesons}, they form the particles called \emph{hadrons}, which obey the strong nuclear force.}, {\it protons} (the number of protons defines the charge number Z) and {\it neutrons} (the sum of the numbers of protons and neutrons defines the mass number A), written as $^A$X for an element 'X'. Note that some isotopes may exist in different nuclear quantum states which have significant stability by themselves, so that transitions between these configurations may liberate the binding energy differences; such states of the same isotope are called {\it isomers}. The landscape of isotopes is illustrated in Fig.~\ref{fig_1_table-of-isotopes}, with black symbols as the naturally-existing stable isotopes, and coloured symbols for unstable isotopes.

Unstable isotopes, once produced, will be \emph{radioactive}, i.e. they will transmute to other isotopes through nuclear interactions, until at the end of such a decay chain a stable isotope is produced. 
Weak interactions will mediate transitions between protons and neutrons and lead to neutrino emission, involvements of atomic-shell electrons 
will result in X-rays from atomic-shell transitions after electron capture and internal-conversion transitions, 
and $\gamma$-rays will be emitted in electromagnetic transitions between excitation levels of a nucleus. 

The production of non-natural isotopes and thus the generation of man-made radioactivity led to the Nobel Prize in Chemistry being awarded to Jean Fr\'ed\'eric Joliot-Curie and his wife Ir\'ene in 1935 -- the second Nobel Prize awarded for the subject of radioactivity after the 1903 award jointly to Pierre Curie, Marie Sk\l odowska Curie, and Henri Becquerel, also in the field of Chemistry.
At the time of writing, element 118 called oganesson (Og) is the most massive superheavy element which has been synthesised and found to exist at least for short time intervals, although more massive elements may exist in an island of stability beyond.  

\noindent Depending on the astrophysical objective, radioactive isotopes may be called \emph{short-lived}, or \emph{long-lived}, depending on how the radioactive lifetime compares to astrophysical time scales of interest. Examples are the utilisation of \Al and \Fe ($\tau\sim$My) diagnostics of the early solar system (\emph{short-lived}, Chap.~6) or of nucleosynthesis source types (\emph{long-lived}, Chap.~3-5).

Which radioactive decays are to be expected? What are stable configurations of nucleons inside the nuclei involved in a production and decay reaction chain? The answer to this involves an understanding of the nuclear forces and reactions, and the structure of nuclei. This is an area of current research, characterised by combinations of empirical modeling, with some capability of \emph{ab initio} physical descriptions, and far from being fully understood. 

Nevertheless, a few general ideas appear well established. 
One of these is recognising a system's trend towards minimising its total energy, and inspecting herein the concept of \emph{nuclear binding energy}. It can be summarised in the expression for nuclear masses \index{mass!nuclear} \citep{Weizsacker:1935}:

\begin{equation}
 m(Z,A) = Z m_p + (A-Z) m_n - BE 
\end{equation}
with
\begin{equation}
 BE = a_{volume} A - a_{surface} A^{2/3} - a_{coulomb} {Z^2 \over {A^{1/3}}} - a_{asymmetry} {{{(a-2Z)}^2} \over {4A}} - {\delta \over A^{1/2}} 
\end{equation}
The total \emph{binding energy} (BE) \index{binding energy} is used as a key parameter for a system of nucleons, and nucleons may thus adopt bound states of lower energy than the sum of the free nucleons, towards a global minimum of system energy. Thus, in a thermal mixture of nucleons, bound nuclei will be formed, and their abundance depends on their composition and shape, and on the overall system temperature, defining how the totally-available phase space of internal and kinetic energy states is populated.
The nucleonic systems would thus have local maxima of binding energy from (1) the \emph{odd-even} \index{odd-even effect} effect described by the last term, which results in odd-nucleon nuclei being less favored that even-nucleon nuclei, and (2) a general excess of neutrons would be favored by the asymmetry term, which results in heavier nuclei being relatively more neutron rich.

The other concept makes use of \emph{entropy}, recognising the relation of this thermodynamic variable to the over-all state of a complex multi-particle and multi-state system. A change in entropy corresponds to a change in the micro-states available to the system. For an infinitesimal change in entropy, we have 
\begin{equation}
T ds = - \sum_i{\mu_idY_i}
\end{equation}
where $Y_i$ are the fractional abundances by number of a species $i$, e.g. $i$= $^{12}$C, or $^4$He, or protons $^1$H, and $\mu$ is the thermodynamic potential\footnote{This is often called \emph{chemical potential}, and describes the energy that is held as internal energy in species $i$, which could potentially be liberated when binding energy per nucleon would change as nucleons would be transferred to different species $j,k,l...$.} of species $i$. 
 
Hence, for our application, if the isotopic composition of a nucleonic mixture changes, its entropy will also change. Or, conversely, the entropy, normalised by the number of baryons in the system, will be a characteristic for the composition:
\begin{equation}
Y_i \propto {{S}\over {n_b}} = s
\end{equation}
with the interpretation of entropy related to the (logarithm of) the number $\Gamma$ of micro-states available:
\begin{equation}
S=k_b \cdot ln \Gamma
\end{equation}
This thermodynamic view allows to calculate \emph{equilibrium} compositions, as they depend on the temperature and on the entropy per baryon. 
With 
\begin{equation}
{{S}\over {n_b}} \propto {{n_\gamma}\over{n_b}}
\end{equation}
the photon to baryon ratio also serves as a measure of the entropy per baryon.
This consideration of thermodynamic equilibrium can be carried through to write down the \emph{nuclear Saha equation} for the composition for an isotope with mass $A$ and charge $Z$:
\begin{eqnarray}
Y_i=Y(Z_i,A_i)=G(Z_i,A_i) [\zeta(3)^{A_i-1} \pi^{(1-A_i)/2} 2^{(3A_i-5)/2} \cdot  \nonumber \\  A_i^{3/2} (k_B T / m_N c^2)^{3(A_i-1)/2}
	 \Phi^{1-A_i} Y_p^{Z_i} Y_n^{A_i-Z_i} exp[BE(Z_i,A_i)/ k_B T ] 
\end{eqnarray}
Herein, $G(Z_i,A_i)$ is the nuclear partition function giving the number of micro-states for the particular isotope, $\zeta(3)$ is the Riemann function of argument 3, and we find again the \emph{binding energy} $BE$ and also the \emph{thermal energy} $k_BT$. $\Phi$ is defined as ratio of photon number to baryon number, and is proportional to the entropy per baryon, thus including the phase space for the plasma constituents. This equation links the proton and neutron abundances to the abundances of all other isotopes, with the characteristic isotope properties of mass $m_N$, mass and charge numbers $A,Z$, and internal micro-states $G$, using the different forms of energy (rest mass, thermal, and binding), as well as the characteristic entropy.  
  
Illustrative examples of how entropy helps to characterise isotopic mixtures are: For high temperatures and entropies, a composition with many nuclei, such as rich in $\alpha$ nuclei would be preferred (e.g. near the big bang in the early universe), while at lower entropy values characteristic for stellar cores a composition of fewer components favouring tightly-bound nucleons in Fe nuclei would be preferred (e.g. in supernova explosions).

With such knowledge about nuclear structure in hand, we can look at the possible configurations that may exist: Those with a minimum of total energy will be \emph{stable}, all others \emph{unstable} or \emph{radioactive}.
 \index{isotopes!table of} Fig.~\ref{fig_1_table-of-isotopes} shows the table of isotopes, encoded as stable (black) and unstable isotopes, the latter decaying by $\beta^-$-decay (blue) and $\beta^+$-decay (orange).
 This is an illustration of the general patterns among the available nuclear configurations.
The \emph{ragged} structure signifies that there are systematic variations of nuclear stability with nucleon number, some nucleonic numbers allowing for a greater variety of stable configurations of higher binding energy. These are, in particular, {\it magic numbers} of protons and neutrons of 2, 8, 20, 28, 50, and 82.
We now know approximately 3100 such \emph{isotopes} making up the 118 now-known \index{elements!known} chemical elements, but only 286 of these isotopes are considered stable. The (7$^{th}$) edition of the Karlsruher Nuklidkarte (2007) \citep{2007KNucChart..7} lists 2962 experimentally-observed isotopes and 652 isomers, its first edition (1958) included 1297 known isotopes of 102 then-known \index{elements!superheavy} elements. 
Theoretical models of atomic nuclei, on the other hand, provide estimates of what might still be open to discovery, in terms of isotopes that might exist but either were not produced in the nearby universe or are too shortlived to be observed. Recent models predict existence of over 9000 nuclei \citep{Erler:2012,Xia:2018}.

It is the subject of this book to explain in detail the astrophysical implications of this characteristic process of nuclear rearrangements, and what can be learned from measurements of the messengers of radioactive decays.
But first we describe the phenomenon of radioactivity in more detail.

\section{Processes of Radioactivity}
\label{sec:1_processes}
  The number of decays at each time should be proportional to the number of currently-existing radioisotopes:  
\begin{equation}
 {{dN}\over{dt}} =  -\lambda \cdot N 
\end{equation}
Here $N$ is the number of isotopes, and the {\it radioactive-decay constant} $\lambda$ \index{decay!constant} is the characteristic of a particular radioactive species.

Therefore, in an ensemble consisting of a large number of identical and unstable isotopes, their number remaining after radioactive decay  \index{decay!radioactive} declines exponentially with time:
\begin{equation}
\label{eq_1} \index{decay!exponential}
 N = N_0 \cdot exp{-t\over\tau} 
\end{equation}
The decay time $\tau$ is the inverse of the radioactive-decay constant\index{decay!time}, and  $\tau$ characterises the time after which the number of isotopes is reduced by decay to $1/e$ of the original number. Correspondingly, the radioactive half-life $T_{1/2}$, \index{decay!half life} is defined as the time after which the number of isotopes is reduced by decay to $1/2$ of the original amount, with
\begin{equation}
   T_{1/2} = {\tau \over ln(2)} 
\end{equation}

The above exponential decay law is a consequence of a surprisingly simple physical property: The probability per unit time for a single radioactive nucleus to decay is independent of the age of that nucleus. Unlike our common-sense experience with living things, decay does not become more likely as the nucleus ages. 
Radioactive decay is a nuclear transition from one set of nucleons constituting a nucleus to a different and energetically-favored set with the same number of nucleons. Different types of interactions can mediate such a transition (see below). In \emph{$\beta$-decays} it is mediated by the \emph{weak transition} \index{weak interaction} of a  neutron into a proton and vice versa\footnote{The mass of the neutron exceeds that of the proton by 1.2933 MeV, making the proton the most stable baryon}, or more generally, nucleons of one type into the other type\footnote{In a broader sense, nuclear physics may be considered to be similar to chemistry: elementary building blocks are rearranged to form different species, with macroscopically-emerging properties such as, e.g., characteristic and well-defined energies released in such transitions.}:
\begin{equation}\label{eq_n-decay}
n \longrightarrow p \mbox{ } + e^- \mbox{ } + \overline{\nu_e} 
\end{equation}
\begin{equation}\label{eq_p-decay}
p \longrightarrow n \mbox{ } + e^+ \mbox{ } + {\nu_e}
\end{equation}
If such a process occurs inside an atomic nucleus, the quantum state of the nucleus is altered. Depending on the variety of configurations in which this new state may be realized (i.e. the \emph{phase space} available to the decaying nucleus), this change may be more or less likely, in nature's attempt to minimize the total energy of a composite system of nucleons. 
The decay probability $\lambda$ per unit time for a single radioactive nucleus is therefore a property which is specific to each particular type of isotope. It can be estimated by Fermi's \emph{Golden Rule} formula \index{Golden Rule} though time-dependent perturbation theory \citep[e.g.][]{1962qume.book.....M}. When schematically simplified to convey the main ingredients, the decay probability is:
\begin{equation}\label{eq1}
\lambda = \frac{4\pi^2}{h} \mbox{ } V_{fi}^2 \mbox{ } \rho(W) 
\end{equation}
\noindent where $\rho(W)$ is the number of final states having suitable energy $W$. The detailed theoretical description involves an integral over the final kinematic states, suppressed here for simplicity. The matrix element $V_{fi}$ is the result of the transition-causing potential between initial and final states.  

In the general laboratory situation, radioactive decay involves a transition from the ground state of the parent nucleus to the daughter nucleus in an excited state. But in cosmic environments, nuclei may be part of hot plasma, and temperatures exceeding millions of degrees lead to population of excited states of nuclei. Thus, quantum mechanical transition rules may allow and even prefer other initial and final states, and the nuclear reactions involving a radioactive decay become more complex. Excess binding energy will be transferred to the end products, which are the daughter nucleus plus emitted (or absorbed, in the case of electron capture transitions) leptons (electrons, positrons, neutrinos) and $\gamma$-ray photons. 

The occupancy of nuclear states is mediated by the \emph{thermal} excitation spectrum of the \emph{Boltzmann distribution} \index{Boltzmann distribution} of particles, populating states at different energies according to:
\begin{equation}
{dN \over dE} = G_j \cdot e^{-{{E}\over{k_BT}}} 
\end{equation}
\noindent Here $k_B$ is Boltzmann's constant, $T$ the temperature of the particle population, $E$ the energy, and $G_j$ the statistical weight factor of all different possible states $j$ which correspond to a specific energy $E$\footnote{States may differ in their quantum numbers, such as spin, or orbital-momenta projections; if they obtain the same energy $E$, they are called \emph{degenerate}.}.
In natural environments, particles will populate different states as temperature dictates. Transition rates among states thus will depend on temperature. Inside stars, and more so in explosive environments, temperatures can reach ranges which are typical for nuclear energy-level differences. Therefore, in cosmic sites, radioactive decay time scales may be significantly different from what we measure in terrestrial laboratories on \emph{cold} samples (see Section~\ref{sec:1_processes} for more detail).

Also the atomic-shell environment of a nucleus may modify radioactive decay,  if a decay involves \emph{capture or emission of an electron} \index{decay!electron capture} to transform a proton into a neutron, or vice versa. Electron capture decays are inhibited in fully-ionized plasma, due to the non-availability of electrons. Also $\beta^-$-decays are affected, as the phase space for electrons close to the nucleus is influenced by the population of electron states in the atomic shell.
  
After Becquerel's discovery of radioactivity \index{radioactivity}  in 1896, Rutherford \index{Rutherford, E.} and others found out in the early 20$^{\rm{th}}$ century that there were different types of radioactive decay \citep{1903PPSL...18..595R}. They called them \emph{$\alpha$ decay, $\beta$ decay} and \emph{$\gamma$ decay}, terms which are still used today. It was soon understood that they are different types of interactions, all causing the same, spontaneous, and time-independent decay of an unstable nucleus into another and more stable nucleus. 

 \noindent {\it Alpha decay}\index{alpha decay}\index{decay!alpha decay}  : This describes the ejection of a $^4$He nucleus from the parent radioactive nucleus upon decay. $^4$He nuclei have since been known also as \emph{alpha particles} for that reason. This decay is intrinsically fast, as it is caused by the \emph{strong} nuclear interaction quickly clustering the nucleus into an alpha particle and the daughter nucleus. Since $\alpha$-nuclei are tighly-bound, they have been found as sub-structures even within nuclei. In the cases of nuclei much heavier than Fe, a nucleus thus consisting of many nucleons and embedded $\alpha$ clusters can find a preferred state for its number of nucleons by separation of such an $\alpha$ cluster, liberating the binding-energy difference\footnote{The binding energy \emph{per nucleon} is maximized for nucleons bound as a Fe nucleus.}. In such heavy nuclei, Coulomb repulsion helps to overcome the potential barrier which is set up by the strong nuclear force, and decay can occur through emission of an $\alpha$ particle. The $\alpha$ particle \emph{tunnels}, with some calculable probability, through the potential barrier, towards an overall more stable and less-energetic assembly of the nucleons.

An example of $\alpha$ decay is $_{88}$Ra$^{226}$ $\Rightarrow$ $_{86}$Rn$^{222}$ + $_2$He$^4$, which is one step in the decay series starting from $^{238}$U. The daughter nucleus , $_{86}$Rn$^{222}$, has charge $Z-2$, where $Z$ is the original charge of the radioactive nucleus ($Z$=88 in this example), because the $\alpha$ particle carried away two charge units from the original radioactive nucleus. Such decay  frequently leads to an excited state of the daughter nucleus. Kinetic energy $E_{\alpha}$ for the $\alpha$ particle is made available from the nuclear binding energy liberation expressed by the \emph{Q-value} of the reaction \index{Q value} if the mass of the radioactive nucleus exceeds the sum of the masses of the daughter nucleus and of the helium nucleus\footnote{These masses may be either nuclear masses or atomic masses, the electron number is conserved, and their binding energies are negligible, in comparison.}:
\begin{equation} \index{isotopes!226Ra}
Q_{\alpha} = [M(_{88}\rm{Ra}^{226}) - M(_{86}\rm{Rn}^{222}) - M(_2\rm{He}^4)]\rm{c}^2
\end{equation}
The range of the $\alpha$ particle (its stopping length) is about 2.7 cm in standard air (for an $\alpha$ particle with E$_{\alpha}$ of 4 MeV), and it will produce about 2$\times$10$^5$ ionizations before being stopped. Even in a molecular cloud, though  its range would be perhaps 10$^{14}$ times larger, the $\alpha$ particle would not escape from the cloud. Within small solids (dust grains), the trapping of radioactive energy from $\alpha$ decay provides a source of heat which may result in characteristic melting signatures\footnote{Within an FeNi meteorite, e.g., an $\alpha$ particle from radioactivity has a range of only $\sim$10~$\mu$m.}.

\noindent {\it Beta decay:}\index{beta decay} \index{decay!beta decay}  This is the most-peculiar radioactive decay type, as it is caused by the nuclear \emph{weak interaction} which converts neutrons into protons and vice versa. The neutrino $\nu$ \index{neutrino} carries energy and momentum to balance the dynamic quantities, as Pauli famously proposed in 1930 (Pauli did not publish this conjecture until 1961 in a letter he wrote to colleagues). The $\nu$ was given its name by \index{Fermi, E.} Fermi, and was discovered experimentally in 1932 by \index{Chadwick, J.} James Chadwick, i.e. \emph{after} Wolfgang Pauli \index{Pauli, W.} had predicted its existence. Neutrinos from the Sun have been discovered to \emph{oscillate} between flavors. $\beta$ decays are being studied in great detail by modern physics experiments, to understand the nature and mass of the $\nu$. Understanding $\beta$ decay challenges our mind, as it involves several such unfamiliar concepts and particles.

There are three types\footnote{We ignore here two additional $\beta$ decays which are possible from $\nu$ and $\overline{\nu}$ captures, due to their small probabilities.} of $\beta$-decay: 
\begin{equation}\label{eq_beta+-decay}
^A_ZX_N\mbox{ }  \longrightarrow \mbox{ } ^A_{Z-1}X_{N+1} \mbox{ } + e^+ \mbox{ } + \nu_e
\end{equation}
\begin{equation}\label{eq_beta--decay}
^A_ZX_N \mbox{ } \longrightarrow \mbox{ } ^A_{Z+1}X_{N-1} \mbox{ } + e^- \mbox{ } + \overline{\nu_e}
\end{equation}
\begin{equation}\label{eq_beta--decay}
^A_ZX_N \mbox{ } + e^- \longrightarrow \mbox{ } ^A_{Z-1}X_{N+1}  \mbox{ } + {\nu_e}
\end{equation}
In addition to eq.~\ref{eq_n-decay} (\emph{$\beta^-$~decay}), these are the conversion of a proton into a neutron (\emph{$\beta^+$~decay}), and \emph{electron capture}.
The weak interaction itself involves two different aspects with intrinsic and different strength, the  vector and axial-vector couplings. The $V_{fi}^2$ term in eq.~\ref{eq1} thus is composed of two terms. These result in  \index{Fermi!transition}\index{Gamow Teller transition} \emph{Fermi} and \emph{Gamow-Teller transitions}, respectively \citep[see][for a review of weak-interaction physics in nuclear astrophysics]{2003RvMP...75..819L}.

An example of $\beta$ decay is \index{isotopes!13N} $_7^{13}$N $\longrightarrow \mbox{ }_6^{13}$C~+~e$^{+}$~$+$~$\nu$, having mean lifetime $\tau$ near 10 minutes. The kinetic energy $Q$ of the two leptons, as well as the created electron's mass, must be provided by the radioactive nucleus having greater mass than the sum of the masses of the daughter nucleus and of an electron (neglecting the comparatively-small neutrino mass).
\begin{equation}
Q_{\beta} =[M(_7^{13}\rm{N}) - M(_6^{13}\rm{C})- m_{e}]c^2
\end{equation}
\noindent where these masses are nuclear masses, not atomic masses. A small fraction of the energy release $Q_{\beta}$ appears as the recoil kinetic energy of the daughter nucleus, but the remainder appears as the kinetic energy of electron and of neutrino.

\begin{figure}
\centerline{
  \includegraphics[width=0.8\textwidth]{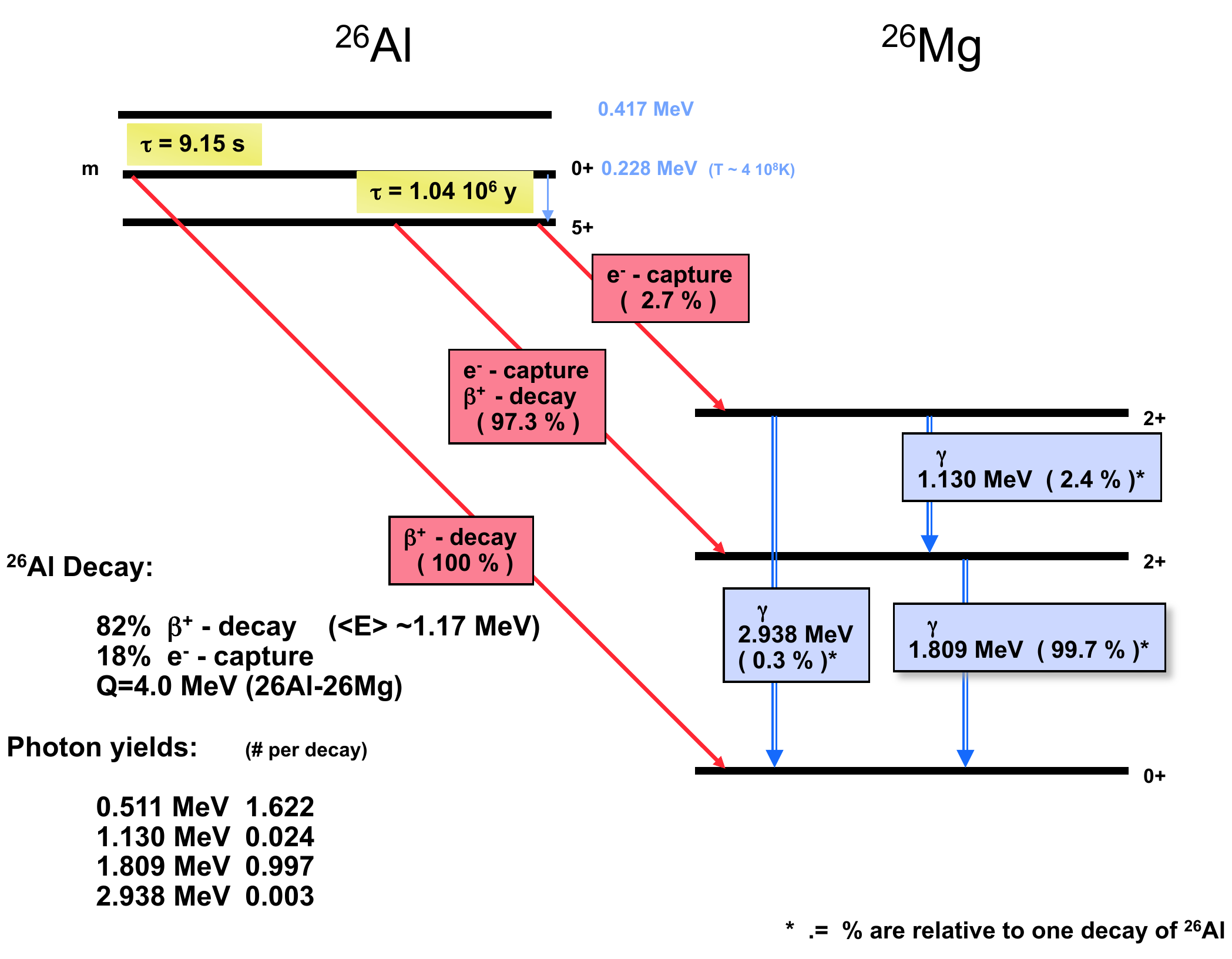}}
  \caption{\Al decay. The \Al nucleus ground state has a long radioactive lifetime, due to the large spin difference of its state to lower-lying states of the daughter nucleus $^{26}$Mg. \index{isotopes!26Al} An isomeric excited state of \Al exists at 228 keV excitation energy. If thermally excited, \Al may decay through this state. Secondary products, lifetime, and radioactive energy available for deposits and observation depend on the environment. }
  \label{fig_1_26Al-decay}
\end{figure}

Capture of an electron is a \emph{two-particle} reaction, the bound atomic electron $e^{-}$ or a free electron in hot plasma being required for this type of $\beta$ decay. Therefore, depending on availability of the electron, electron-capture $\beta$ decay lifetimes can be very different for different environments. 
	In the laboratory case, electron capture usually involves the 1s electrons of the atomic structure surrounding the radioactive nucleus, because those present their largest density at the nucleus. 
	
	In many cases the electron capture competes with $e^{+}$ $+$ $\nu$ emission. In above example, $^{13}$N can decay not only by emitting $e^{+}$ $+$ $\nu$, but also by capturing an electron: $_7^{13}$N~+~e$^{-}\longrightarrow_6^{13}$C~+~$\nu$. In this case the capture of a 1s electron happens to be much slower than the rate of e$^+$ emission. But cases exist for which the mass excess is not large enough to provide for the creation of the $e^{+}$ mass for emission, so that only electron capture remains to the unstable nucleus to decay. Another relevant example is the decay of $^7$Be. Its mass excess over the daugther nucleus $^7$Li is only 0.351 MeV. This excess is insufficient to provide for creation of the rest mass of an emitted $e^{+}$, which is 0.511 MeV. Therefore, the $^7$Be nucleus \index{isotopes!7Be} is stable against $e^{+}$ $+$ $\nu$ emission. However, electron capture adds 0.511 MeV of rest-mass energy to the mass of the $^7$Be nucleus, giving a total 0.862 MeV of energy above the mass of the $^7$Li nucleus. Therefore, the $e^{-}$ capture process (above) emits a monoenergetic neutrino having $E_{\nu}$= 0.862 MeV\footnote{This neutrino line has just recently been detected by the Borexino collaboration arriving from the center of the Sun \citep{2008PhRvL.101i1302A}.}.

	The situation for electron capture processes differs significantly in the interiors of stars and supernovae: Nuclei are ionized in plasma at such high temperature. The capture lifetime of $^7$Be, for example, which is 53 days against 1s electron capture in the laboratory, is lengthened to about 4 months at the solar center \citep[see theory by][]{1964ApJ...139..318B,1983NuPhA.404..578T}, where the free electron density is less at the nucleus.

	The range of the $\beta$ particle (its stopping length) in normal terrestrial materials is small, being a charged particle which undergoes Coulomb scattering. An MeV electron has a range of several meters in standard air, during which it loses energy by ionisations and inelastic scattering. In tenuous cosmic plasma such as in supernova remnants, or in interstellar gas, such collisions, however, become rare, and may be unimportant compared to electromagnetic interactions of the magnetic field (\emph{collisionless plasma}).
Energy deposit or escape is a major issue in intermediate cases, such as expanding envelopes of stellar explosions, supernovae (positrons \index{positron} from $^{56}$Co \index{isotopes!56Co}\index{isotopes!44Ti} and $^{44}$Ti) and novae (many $\beta^+$ decays such as $^{13}$N) (see Chapters~4, 5, and~7 for a discussion of the various astrophysical implications). 
Even in small solids and dust grains, energy deposition from \Al  $\beta$-decay, for example, injects 0.355~W~kg$^{-1}$ of heat. This is sufficient to result in melting signatures, which have been used to study condensation sequences of solids in the early solar system (see Chapter~6).

\noindent {\it Gamma decay:}\index{gamma decay} \index{decay!gamma decay}   In $\gamma$ decay the radioactive transition to a different and more stable nucleus is mediated by the \emph{electromagnetic interaction}. A nucleus relaxes from its excited configuration of the nucleons to a lower-lying state of the same nucleons. This is intrinsically a fast process; typical lifetimes for excited states of an atomic nucleus are 10$^{-9}$seconds. We denote such electromagnetic transitions of an excited nucleus \emph{radioactive $\gamma$-decay} when the decay time of the excited nucleus is considerably longer and that nucleus thus may be considered a temporarily-stable configuration of its own, a \emph{metastable} nucleus. 

How is stability, or instability, of a nuclear-excited state effected?
In electromagnetic transitions 
\begin{equation}\label{eq_photon-decay}
A^{\star} \longrightarrow A^{g.s.} + \gamma
\end{equation}
the spin (angular momentum) is a conserved quantity of the system. The spin of a nuclear state is a property of the nucleus as a whole, and reflects how the states of protons and neutrons are distributed over the quantum-mechanically allowed \emph{shells} or nucleon wave functions (as expressed in the \emph{shell model} view of an atomic nucleus).
The photon (\emph{$\gamma$ quantum}) emitted (eq.\ref{eq_photon-decay}) will thus have a \emph{multipolarity} \index{multipolarity} resulting from the spin differences of initial and final states of the nucleus. Dipole radiation is most common and has multipolarity 1, emitted when initial and final state have angular momentum difference $\Delta l=1$. Quadrupole radiation (multipolarity 2, from $\Delta l=2$) is $\sim$6 orders of magnitude more difficult to obtain, and likewise, higher multipolarity transitions are becoming less likely by the similar probability decreases 
(the \emph{Weisskopf estimates} \index{Weisskopf estimates} \citep[see][]{1951PhRv...83.1073W}). 
This explains why some excited states in atomic nuclei are much more long-lived (\emph{meta-stable}) \index{metastable} than others; their transitions to the ground state are also considered as \emph{radioactivity}, and called \emph{$\gamma$ decay}.

The range of a $\gamma$-ray (its stopping length) is typically about 5-10~g~cm$^{-2}$ in passing through matter of all types. Hence, except for dense stars and their explosions, radioactive energy from $\gamma$ decay is of astronomical implication only\footnote{Gamma-rays from nuclear transitions following $^{56}$Ni decay \index{isotopes!56Ni} (though this is a $\beta$ decay by itself) inject radioactive energy through $\gamma$-rays from such nuclear transitions into the supernova \index{supernova!light curve} envelope, where it is absorbed in scattering collisions and thermalized. This heats the envelope such that thermal and optically bright supernova light is created. Deposition of $\gamma$-rays from nuclear transitions are the engines which make supernovae to be bright light sources out to the distant universe, used in cosmological studies \citep{2000A&ARv..10..179L} to, e.g., support evidence for \emph{dark energy}.}. 

An illustrative example of radioactive decay is the \Al nucleus. \index{decay!26Al} \index{isotopes!26Al}\index{isotopes!26Mg} Its decay scheme is illustrated in Fig.~\ref{fig_1_26Al-decay}. The ground state of \Al is a $5+$ state. Lower-lying states of the neighboring isotope $^{26}$Mg have states $2+$ and $0+$, so that a rather large change of angular momentum $\Delta l$ must be carried by radioactive-decay secondaries. This explains the large $\beta$-decay lifetime of \Al of $\tau\sim$1.04~10$^6$~y.  
In the level scheme of \Al, excited states exist at energies 228, 417, and 1058~keV. The $0+$ and $3+$ states of these next excited states are more favorable for decay due to their smaller angular momentum differences to the $^{26}$Mg states, although $\Delta l=0$ would not be \emph{allowed} for the 228~keV state to decay to $^{26}$Mg's ground state. This explains its relatively long lifetime of 9.15~s, and it is a \emph{metastable} \index{metastable} state of \Al.  If thermally excited, which would occur in nucleosynthesis sites exceeding a few 10$^8$K, \Al may decay through this state without $\gamma$-ray emission as $^{26}\rm{Al}^{g.s.} + \gamma \longrightarrow ^{26}\rm{Al}^{m} \longrightarrow ^{26}\rm{Mg} + e^+$, while the ground state decay is predominantly a \emph{$\beta^+$ decay} through excited $^{26}$Mg states and thus including $\gamma$-ray emission. Secondary products, lifetime, and radioactive energy available for deposits and observation depend on the environment.

\begin{figure}
  \includegraphics[width=\textwidth]{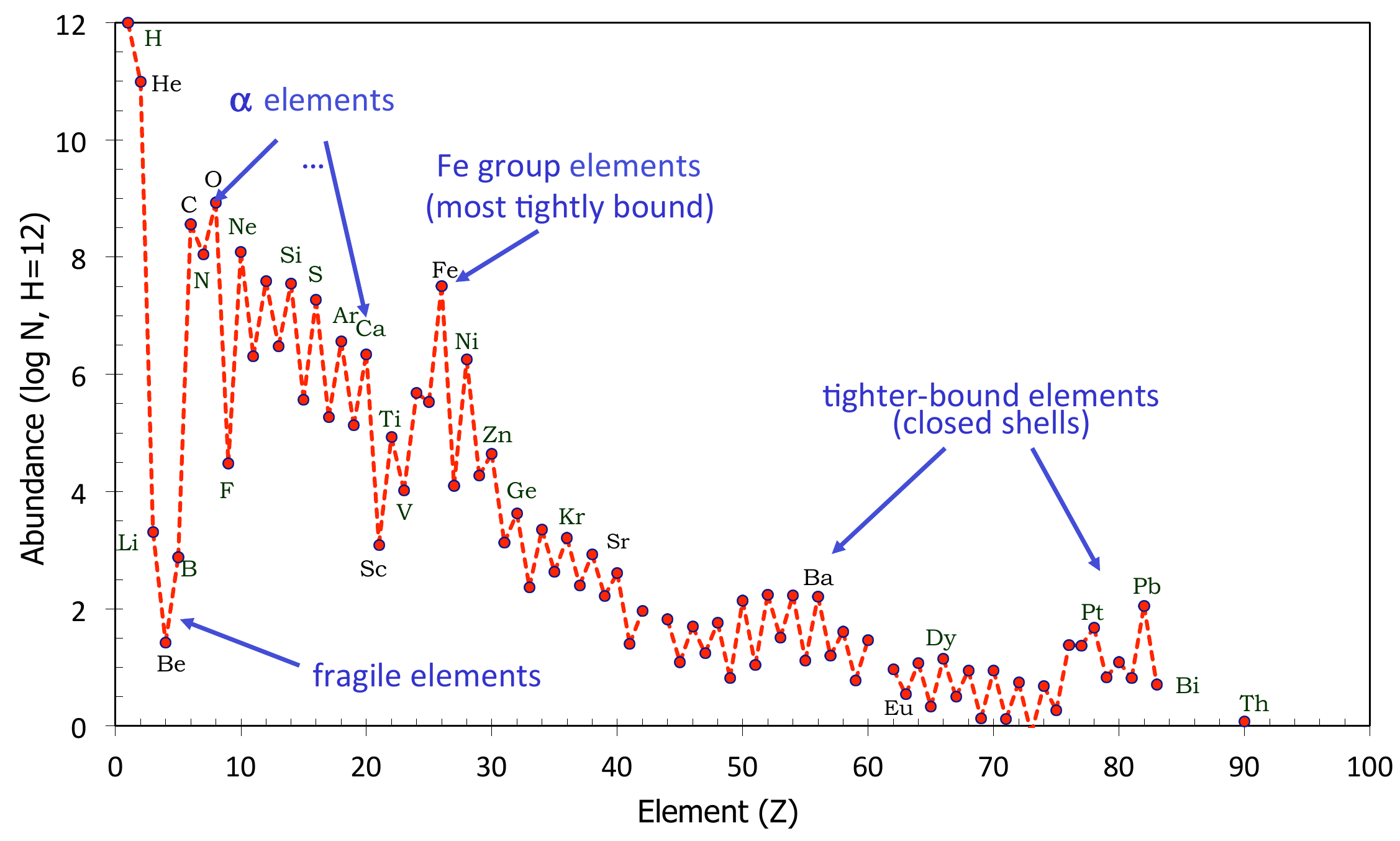}
  \caption{The abundance of elements \index{abundances!cosmic} \index{abundances!elemental}     in the present-day nearby universe. Abundances (by number) are shown in a logarithmic scale, and span 12 orders of magnitude. The interplay of nuclear properties (several are indicated in the graph) with environmental conditions in cosmic nucleosynthesis sites has created this complex abundance pattern during the course of cosmic history.}
  \label{fig_1_abundances}
\end{figure}

\section{Radioactivity and Cosmic Nucleosynthesis}

Nuclear reactions in cosmic sites re-arrange the basic constituents of atomic nuclei (neutrons and protons) among the different allowed configurations.
Throughout cosmic evolution, such reactions occur in various sites with different characteristic environmental properties. Each reaction environment leads to rearrangements of the relative abundances of cosmic nuclei. The cumulative process is called \emph{cosmic chemical evolution}\index{chemical evolution!cosmic}. \footnote{We point out that there is no chemistry involved; the term refers to changes in abundances of the chemical elements, which are important for our daily-life experiences. But these are a result of the more-fundamental changes in abundances of isotopes mediated by cosmic nuclear reactions.}

The \emph{cosmic abundance} of a specific isotope \index{abundances!normalisations} is expressed in different ways, depending on the purpose. Counting the atoms of isotope $i$ per unit volume, one obtains $n_i$, the number density of atoms of species $i$ (atoms~cm$^{-3}$).  The interest of cosmic evolution and nucleosynthesis lies in the fractional abundances of species $i$ related to the total, and how it is altered by cosmic nuclear reactions. Observers count a species $i$ and relate it to the abundance of a reference species. For \index{abundances!astronomical} astronomers this is hydrogen. Hydrogen is the most abundant element throughout the universe, and easily observed through its characteristic atomic transitions in spectroscopic astronomical measurements. 
Using the definition of  Avogadro's constant $A_{Av}$  as the number of atoms which make up $A$ grams of species $i$ (i.e., one mole), we can obtain abundances \emph{by mass}; $A_{Av}=6.02214~10^{23}$ atoms~mole$^{-1}$. 
The mass contained in a particular species $S$ results from scaling its abundance $N_S$ by its atomic weight $A$. 

We can get a global measure for cosmic evolution of the composition of matter by tracing how much of the total mass is contained in hydrogen, helium, and the remainder of elements called \emph{metals}\footnote{This nomenclature may be misleading, it is used by  convenience among astrophysicists. Only a part of these elements are actually metals.}, calling these quantities $X$ for hydrogen abundance, $Y$ for helium abundance, and $Z$ for the cumulative abundance of all nuclei heavier than helium. We call these \emph{mass fractions} of hydrogen $X$, helium $Y$, and metals $Z$, with $X+Y+Z=1$.  
The metalicity $Z$ is a key parameter used to characterise the evolution of elemental and isotopic composition of cosmic matter. The astronomical abundance scale is set from most-abundant cosmic element Hydrogen to $log(X_H)=12$ (Fig.~\ref{fig_1_abundances}), but mineralogists and meteoriticians use $Si$ as their reference element and set $log(X_{Si})=6$. 

We often relate abundances also to our best-known reference, the solar system, denoting \emph{solar-system} values by the $\odot$ symbol. Abundances of a species $S$ are then expressed in \index{abundances!bracket notation} \emph{bracket notation}\footnote{Deviations from the standard may be small, so that $\lbrack\frac{[S_1}{S_2}\rbrack$ may be expressed in $\delta$ units (parts per mil), or $\epsilon$ units (parts in 10$^4$), or ppm and ppb; $\delta(^{29}Si/^{28}Si)$ thus denotes excess of the $^{29}$Si/$^{28}$Si isotopic ratio above solar values in units of 0.1\%.}
 as \index{abundances!cosmic} \index{abundances!solar} \index{abundances!standard} 
\begin{equation}
\lbrack \frac{ S }{ H } \rbrack \equiv log (\frac{X_S}{X_H})_{\star}  -  log (\frac{X_S}{X_H})_{\odot}  
\end{equation}
\noindent
Depending on observational method and precision, our astronomical data are  \emph{metalicity}, elemental \emph{enrichments} with respect to solar abundances, or isotopic abundances. Relations to nuclear reactions are therefore often indirect. Understanding the nuclear processing of matter in the universe is a formidable challenge, often listed as one of the \emph{big questions} of science.

Big Bang Nucleosynthesis (BBN) about 13.8~Gyrs ago \index{abundances!big bang}  \index{nucleosynthesis!big bang} left behind a primordial \index{abundances!primordial} composition where hydrogen (protons) and helium were the most-abundant species; the total amount of nuclei heavier than He (the \emph{metals}) was less than 10$^{-9}$ (by number, relative to hydrogen \citep{2007ARNPS..57..463S}).
Today, the total mass fraction of metals in {\it solar abundances} is $Z=0.0134$  \citep{2009ARA&A..47..481A}, compared to a hydrogen mass fraction of\footnote{This implies a \emph{metalicity} of solar matter of 1.4\%. Our local reference for cosmic material composition seems to be remarkably universal. Earlier than $\sim$2005, the commonly-used value for solar metallicity \index{metallicity!solar} had been 2\%.} $X=0.7381$. This growth of metal abundances by about seven orders of magnitude is the effect of cosmic nucleosynthesis. Nuclear reactions in stars, supernovae, novae, and other places where nuclear reactions may occur, all contribute. But it also is essential that the nuclear-reaction products inside those cosmic objects will eventually be made available to observable cosmic gas and solids, and thus to later-generation stars such as our solar system born 4.6~Gyrs ago. This book will also discuss our observational potential for cosmic isotopes, and we address the constraints and biases which limit our ability to draw far reaching conclusions.

The growth of isotopic and elemental abundances from cosmic nucleosynthesis does not occur homogeneously. Rather, the cosmic abundances observed today span a dynamic range of twelve orders of magnitude between abundant hydrogen and rare heavy elements (Fig.~\ref{fig_1_abundances}). Moreover, the elemental abundance pattern already illustrates clearly the prominent effects of nuclear structure (see Fig.~\ref{fig_1_abundances}): Iron elements are among the most-tightly bound nuclei, and locally elements with even numbers of nucleons are more tightly bound than elements with odd numbers of nuclei. The Helium nucleus (\emph{$\alpha$-particle}) also is more tightly bound than its neighbours in the chart of nuclei, hence all elements which are multiples of $\alpha$'s are more abundant than their neighbours. 

Towards the heavier elements beyond the Fe group, abundances drop by about five orders of magnitude again, signifying a substantially-different production process than the mix of charged-particle nuclear reactions that produced the lighter elements: \emph{neutron capture} on Fe \emph{seed nuclei}. The two abundance peaks seen for heavier elements are the results of different environments for cosmic neutron capture reactions (the \emph{r-process} and \emph{s-process}), both determined by neutron capture probabilities having local extrema near \emph{magic numbers}. The different peaks arise from the particular locations at which the processes' reaction path encounters these \emph{magic nuclei}, as neutron captures proceed much faster (slower) than $\beta$~decays in the $r$~process ($s$~process)..

The subjects of cosmic nucleosynthesis research are complex and diverse, and cover the astrophysics of stars, stellar explosions, nuclear reactions on surfaces of compact stars and in interstellar space. For each of the potential nuclear-reaction sites, we need to understand first how nuclear reactions proceed under the local conditions, and then how material may be ejected into interstellar space from such a source. None of the nucleosynthesis sites is currently understood to a  level of detail which would be sufficient to formulate a physical description, sit back and consider cosmic nucleosynthesis \emph{understood}. For example,  one might assume we know our Sun as the nearest star in most detail; but solar neutrino \index{neutrino} measurements have been a puzzle only alleviated in recent years with the revolutionary adoption of non-zero masses for neutrinos, whih allow for flavour oscillations; and even then, the abundances of the solar photosphere, revised by almost a factor two based on three-dimensional models of the solar photosphere  \citep{Asplund:2009}, created surprising tension with measurements of helio-seismology and the vibrational behaviour reflected herein, and the physical descriptions in our currently-best solar model are under scrutiny \citep{Vinyoles:2017}.
  
As another example, there are two types of supernova explosions. Core-collapse supernovae are the presumed outcome of the final gravitational collapse of a massive star once its nuclear fuel is exhausted, and thermonuclear supernovae were thought to originate from detonation of degenerate stars once they exceed a critical threshold for nuclear burning of Carbon, the Chandrasekhar mass limit. The gravitational collapse can not easily be reverted into an explosion, and even the help of neutrinos from the newly-forming neutron star in the center appears only marginally sufficient, so that many massive stars that were thought to explode may collapse to black holes \citep{Janka:2016}. And the thermonuclear supernova variety appears to require white dwarf collisions as triggering events in some well-constrained cases, while in other cases luminosities deviate by orders of magnitude from the expectation from a Chandrasekhar-mass white dwarf and its nuclear-burning demise that once was thought to be a cosmic standard candle \citep{Hillebrandt:2013}. For neither of these supernovae, a \emph{physical} model is available, which would allow us to calculate and predict the outcome (energy and nuclear ashes)  under given, realistic, initial conditions (see Ch. 4 and 5). 
Much research remains to be done in cosmic nucleosynthesis.

One may consider measurements of cosmic material in all forms to provide a wealth of data, which now has been exploited to understand cosmic nucleosynthesis. Note, however, that cosmic material as observed has gone through a long and ill-determined journey. We need to understand the trajectory in time and space of the progenitors of our observed cosmic-material sample if we want to interpret it in terms of cosmic nucleosynthesis. 
This is a formidable task, necessary for distant cosmic objects, but here averaging assumptions help to simplify studies. For more nearby cosmic objects where detailed data are obtained, astrophysical models quickly become very complex, and also need simplifying assumptions to operate for what they are needed. It is one of the objectives of cosmic nucleosynthesis studies to contribute to proper models for processes in such evolution, which are sufficiently isolated to allow their separate treatment. Nevertheless, carrying out \emph{well-defined experiments} for a source of cosmic nucleosynthesis remains a challenge, due to this complex flow of cosmic matter (see Ch.'s 6 to 8).

The special role of radioactivity in such studies is contributed by the intrinsic decay of such material after it has been produced in cosmic sites. This brings in a new aspect, the clock of the radioactive decay. Technical applications widely known are based on $^{14}$C with its half life of 5700 years, while astrophysical applications extend this to much longer half lives up to Gyrs ($^{235}$U has a decay time of 10$^9$~years). Changes in isotopic abundances with time will occur at such natural and isotope-specific rates, and will leave their imprints in observable isotopic abundance records. For example, the observation of unstable technetium in stellar atmospheres of AGB stars was undisputable proof of synthesis of this element inside the same star, because the evolutionary time of the star exceeds the radioactive lifetime of technetium. Another example, observing radioactive decay $\gamma$-ray lines from short-lived Ni isotopes from a supernova is clear proof of its synthesis in such explosions; measuring its abundance through $\gamma$-ray brightness is a direct \emph{calibration} of processes in the supernova interior. A last example, solar-system meteorites \index{abundances!solar} show enrichments in daughter products of characteristic radioactive decays, such as $^{26}$Al \index{isotopes!26Al}\index{isotopes!53Mn} and $^{53}$Mn; the fact that these radioactive elements were still alive at the time those solids formed sets important constraints to the time interval between the latest nucleosynthesis event near the forming Sun and the actual condensation of solid bodies in the interstellar gas accumulating to form the young solar system. This book will discuss these examples in detail, and illustrate the contributions of radioactivity studies to the subject of cosmic nucleosynthesis.

\section{Observing radioactive Isotopes in the Universe} 

\begin{figure} 
  \includegraphics[width=\textwidth]{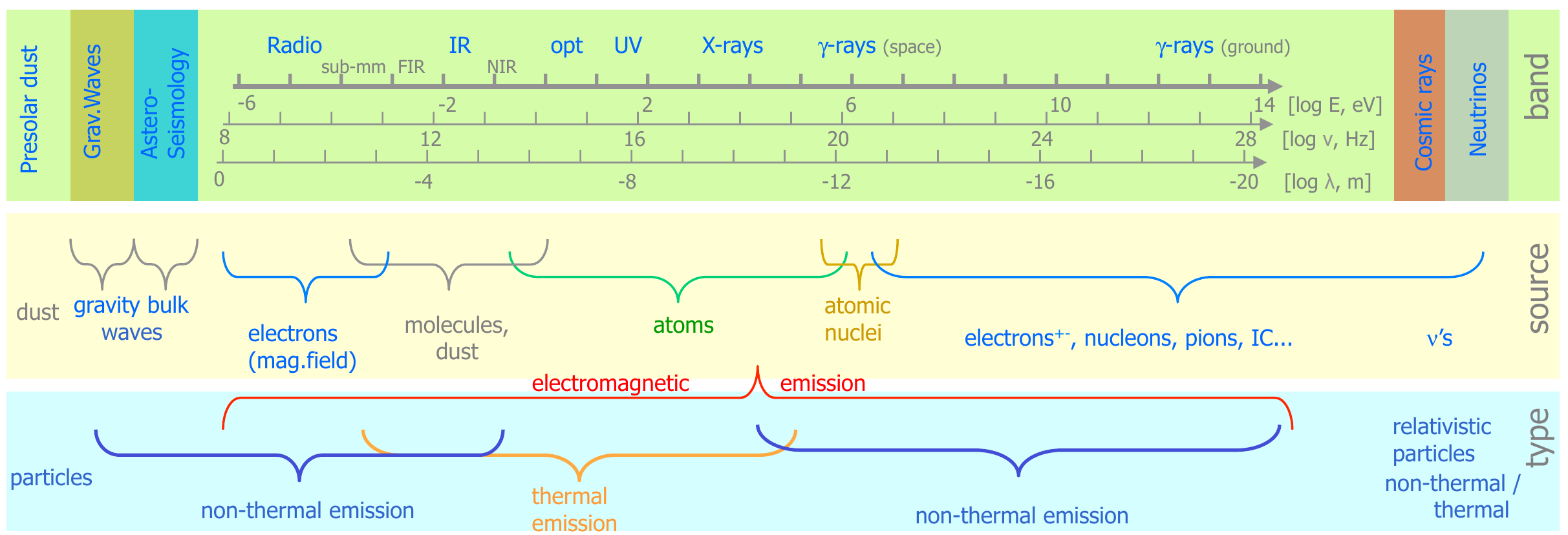}
  \caption{The electromagnetic spectrum \index{spectrum!electromagnetic}\index{electromagnetic radiation}  of candidate astronomical measurements ranges across more than twenty orders of magnitude. Not all are easily accessible. Information categories of thermal and non-thermal, and of molecular, atomic, nuclear, and elementary-particle physics origins of cosmic radiation extends over different parts of this broad spectrum. Nuclear physics is directly accessible in a small band (0.1-10 MeV) only. Non-electromagnetic astronomical messengers are indicated at both ends of the electromagnetic spectrum}
  \label{fig_1_emSpectrum}
\end{figure} 

 Astronomy has expanded beyond the narrow optical band into \emph{new astronomies} \index{astronomy} in the past decades. By now, we are familiar with telescopes measuring radio and sub-mm through infrared emission towards the long wavelength end, and ultraviolet, X-ray, and $\gamma$-ray emission towards the short wavelength end (see Fig.~\ref{fig_1_emSpectrum}). \index{spectrum!electromagnetic} The physical origins of radiation are different in different bands. 
Thermal radiation \index{radiation!thermal} dominates emission from cosmic objects in the middle region of the electromagnetic spectrum, from a few 10K cold molecular clouds at radio wavelengths through dust and stars up to hot interstellar gas radiating X-rays. \index{radiation!non-thermal} Non-thermal emission is characteristic for the wavelength extremes, both at radio and $\gamma$-ray energies. Characteristic spectral lines originate from atomic shell \index{spectrum!atomic}  electrons over most of the spectrum; nuclear lines are visible only in roughly two decades of the spectrum \index{spectrum!nuclear}  at 0.1--10~MeV. 
Few exceptional lines arise at high energy from annihilations \index{spectrum!annihilation} of \index{positron}\index{pion} positrons and pions. Cosmic \emph{elements} can be observed in a wide astronomical range. \emph{Isotopes}, however, are observed almost exclusively through $\sim$MeV $\gamma$-rays (see Fig.~\ref{fig_1_emSpectrum}). Note that nucleosynthesis reactions occur among isotopes, so that this is the prime\footnote{Other astronomical windows may also be significantly influenced by biases from other astrophysical and astrochemical processes; an example is the observation of molecular isotopes of CO, where chemical reactions as well as dust formation can lead to significant alterations of the abundance of specific molecular species.} information of interest when we wish to investigate cosmic nucleosynthesis environment properties.

\begin{figure}
\centerline{
  \includegraphics[width=0.45\textwidth]{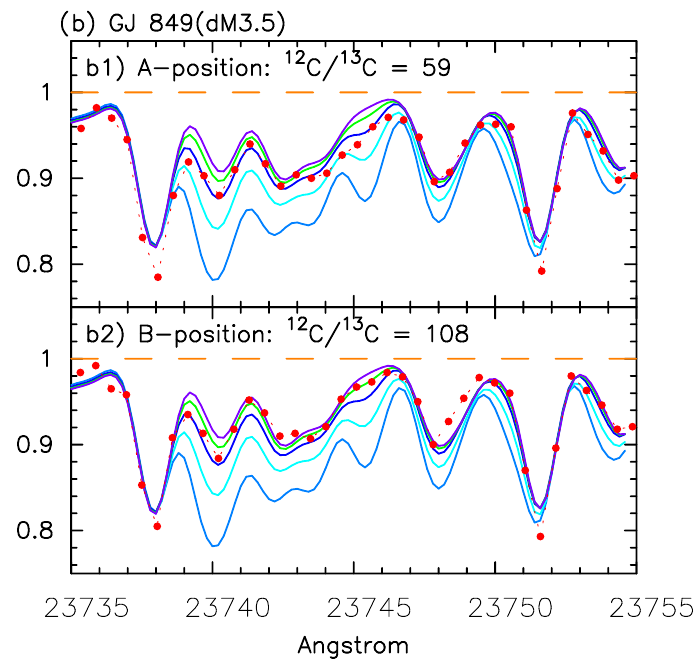}   \includegraphics[width=0.55\textwidth]{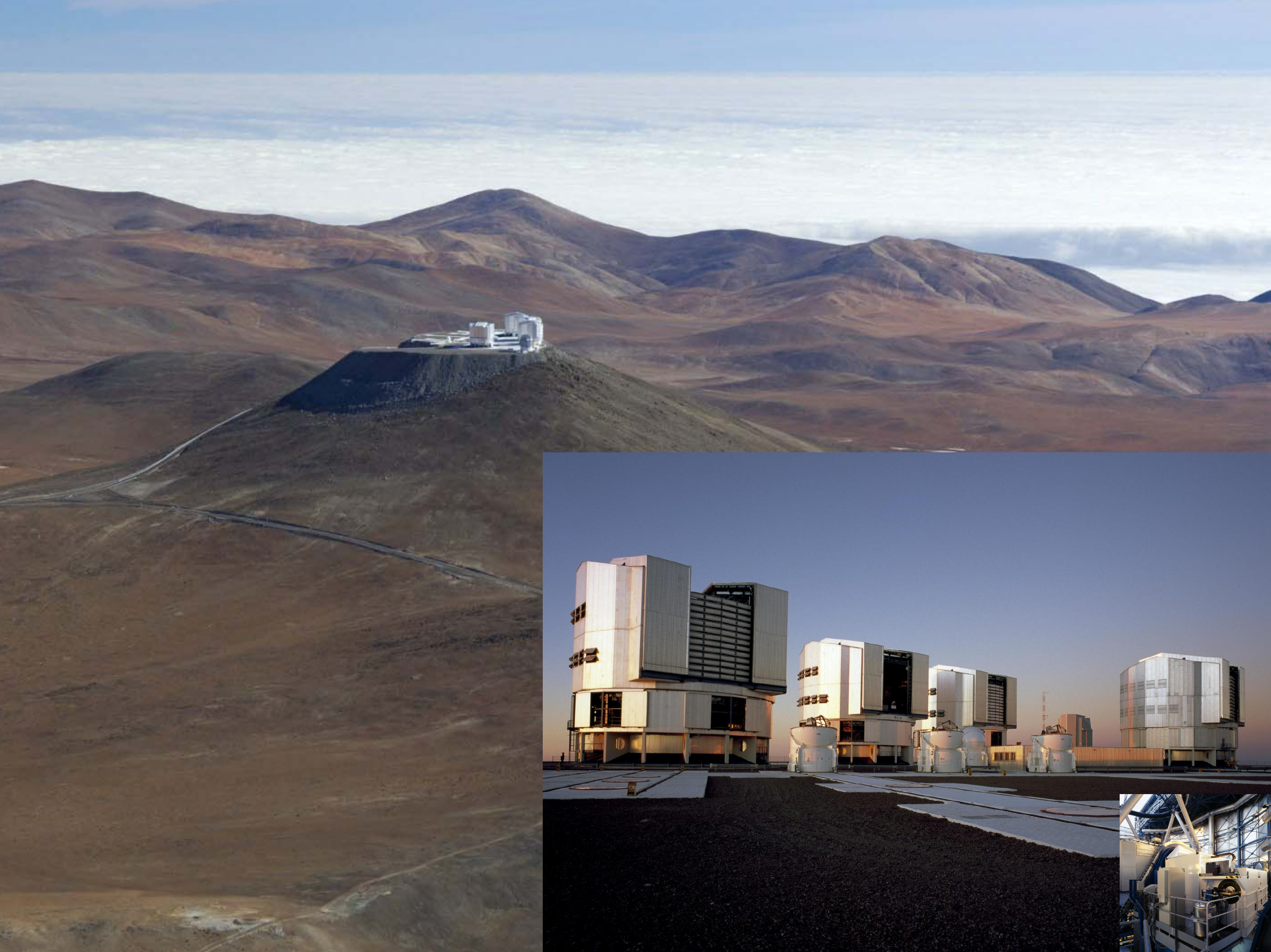} }
  \caption{\emph{Left:} Example of an isotopic measurement in a stellar atmosphere. Shown is an absorption-line spectrum \index{spectrum!absorption lines} of a cool star with a present-generation optical telescope, here the Subaru telescope on Hawaii with its IR spectrograph at a resolution of 20000. Molecular lines from the CO molecule isotopologes show isotopic shifts, which can be recognised as changes in line shapes, as resulting from the isotopic abundance ratio.  Here the carbon isotopic ratio is determined for the stellar atmosphere of a M dwarf star, comparing the measurement (red dots) with expectations for different ratios $^{12}$C/$^{13}$C \citep[from][]{Tsuji:2016}. 
  \emph{Right:} The Very Large Telescope (VLT) on \index{Very Large Telescope (VLT)} Mount Paranal in Chile, with four telescopes (lower right), is one of the modern optical instruments. Equipped with high-resolution spectrographs such as FLAMES (insert lower right), absorption-line spectroscopy of stars  provides elemental abundances in stellar atmospheres, even in nearby galaxies. (Figure ESO) }
  \label{fig_1_stellar_spectrosscopy}
\end{figure}

Only few elements such as technetium (Tc) \index{elements!Tc} \index{isotopes!98Tc} do not have any stable isotope; therefore, elemental photospheric absorption and emission line spectroscopy, the backbone of astronomical studies of cosmic nucleosynthesis, have very limited application in astronomy with radioactivities. This is about to change currently, as spectroscopic devices in the optical and radio/sub-mm regimes advance spectral resolutions. 
Observational studies of cosmic radioactivities are best performed by techniques which intrinsically obtain isotopic information. These are:
\begin{itemize}
\item
Modern spectrographs on large ground-based telescopes reach R=20000, sufficient to resolve fine structure lines and isotopic features in molecules (see Fig.~\ref{fig_1_stellar_spectrosscopy}). Radio spectroscopy with CO isotopes has been successfully applied since the 1990ies, and has been used mainly to track the CO molecule at different columns densities, while sub-mm lines from molecules have been demonstrated to observe specific isotopes within molecules such as $^{36}$ArN \citep{Schilke:2014}. 
\item 
Precision mass spectroscopy \index{mass!spectroscopy} in terrestrial laboratories, which has been combined with sophisticated radiochemistry to extract meteoritic components originating from outside the solar system \index{meteorites}
\item 
Spectroscopy of characteristic $\gamma$-ray lines \index{gamma-ray lines} \index{spectrum!gamma-ray} emitted upon radioactive decay in cosmic environments
\end{itemize}

\noindent The two latter {\it astronomical disciplines} have a relatively young history. They encounter some limitations due to their basic methods of how astronomical information is obtained, which we therefore discuss in somewhat more detail:
\begin{itemize}
\item
Precision mass spectrometry of meteorites \index{meteorites} for astronomy with radioactivity began about 1960 with a new discovery of  now extinct radioactivity within the young solar system.  From heating of samples of bulk meteorite material, the presence of \index{isotopes!129Xe} \index{isotopes!129I} a surprising excess  $^{129}$Xe had been puzzling. Through a variety of different chemical processing, this could be tracked to trapped gas enclosures in rather refractory components, which must have been enriched in $^{129}$I at the time of formation of this meteorite. From mineralogical arguments, this component could be associated with the early solar system epoch about 4.6~Gy ago \citep{PhysRevLett.4.8}. This was the first evidence that the matter from which the solar system formed contained radioactive nuclei whose half-lives are too short to be able to survive from that time until today ($^{129}$I decays to $^{129}$Xe within 1.7~10$^7$y). 
Another component could be identified from most-refractory Carbon-rich material, and was tentatively identified with dust grains of pre-solar origins. Isotopic anomalies found in such \emph{extra-solar} inclusions, e.g. for C and O isotopes, range over four orders of magnitude for such \emph{star dust} grains as shown in Fig.~\ref{fig_1_grain} \index{stardust} \citep{1998AREPS..26..147Z}, while isotopic-composition variations among bulk meteoritic-material samples are a few percent at most. These mass spectroscopy measurements are characterised by an amazing sensitivity and precision, clearly resolving isotopes and counting single atoms at ppb levels to determine isotopic ratios of such rare species with high accuracy, and nowadays even for specific, single dust grains. 
This may be called an \emph{astronomy in terrestrial laboratories} (see Chapter~11 for instrumental and experimental aspects), and is now an established part of astrophysics \citep[see ][for a recent review]{2004ARA&A..42...39C} and \citep[e.g.][]{Amari:2014,Zinner:2014}.  
\begin{table*}
\begin{tabular}{|c|c|c|c|c|}
\hline 
    {\bf  Isotope }& {\bf Lifetime} & {\bf Presolar Grain } & {\bf Source}  & Ref. \\
   {\bf  chain } & & {\bf Type} & {\bf  } & \\
    \hline \hline
   $^{49}$V    $\longrightarrow$    $^{49}$Ti   &       330 days      &        SiC, Graphite      &      SNe     & [1]       \\
  \, $^{22}$Na     $\longrightarrow$    $^{22}$Ne     &    2.6 years      &    Graphite      &    SNe     &  [2]  \\
  \, $^{44}$Ti      $\longrightarrow$    $^{44}$Ca       &    60 years      &    SiC, Graphite, Hibonite      &    SNe     & [3]    \\
  \, $^{32}$Si      $\longrightarrow$    $^{32}$S       &    153 years      &    SiC      &    SNe, post-AGB stars  & [4]       \\
  \, $^{41}$Ca      $\longrightarrow$   $^{41}$K     &    1.02 10$^5$ years      &    SiC, Graphite, Hibonite      &    SNe, RGB, and AGB stars   &  [5]     \\
  \, $^{99}$Tc     $\longrightarrow$   $^{99}$Ru     &    2.11 10$^5$ years      &    SiC      &    AGB stars   & [6]      \\
  \, $^{26}$Al    $\longrightarrow$    $^{26}$Mg       &    7.17 10$^5$ years      &    SiC, Graphite, Corundum,       &    SNe, RGB, and AGB stars    & [7]   \\
  \,       \quad          &           &      Spinel, Hibonite, Silicate  &   &    \\
    \, $^{93}$Zr      $\longrightarrow$   $^{93}$Nb    &    1.61 10$^6$ years      &    SiC      &    AGB stars    & [8]   \\
     \hline
 \end{tabular}
\caption{Radioactivities in presolar grains, sorted by ascending radioactive mean lifetime \citep[from][]{Groopman:2015}. 
References: [1]  Hoppe and Besmehn (2002 [2] Amari (2009)    [3] Nittler et al. (1996)    [4]  Pignatari et al. (2013),Fujiya et al. (2013)    [5] Amari et al. (1996)    [6] Savina et al. (2004)  [7] Zinner et al. (1991)    [8] Kashiv et al. (2010) (see Groopman et al. (2015) for these references).}
\label{fig:TableRadioactGrains}
\end{table*}

\begin{figure}
 \centerline{
  \includegraphics[width=0.6\textwidth]{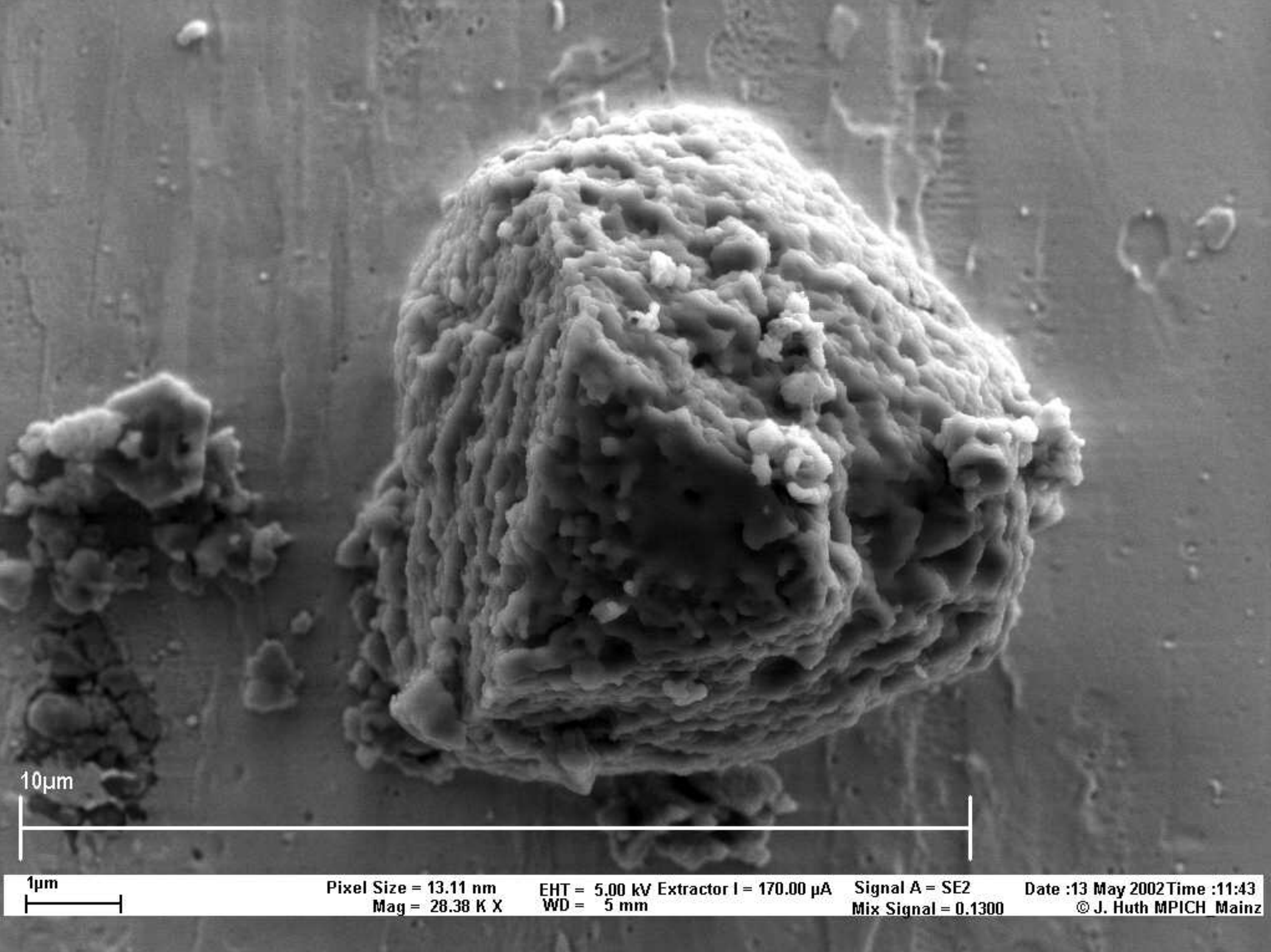}}
  \caption{Meteoritic inclusions such as this SiC grain \index{meteorites} are recognised as dust formed near a cosmic nucleosynthesis source outside the solar system, from their large isotopic anomalies, which cannot be explained by interstellar nor solar-system processing but are reminiscent of cosmic nucleosynthesis sites. Having condensed in the envelope of a source of new isotopes, laboratory mass spectroscopy can reveal isotopic composition for many elements, thus providing a remote probe of one cosmic nucleosynthesis source.}
 \label{fig_1_grain}
\end{figure}

Table 1.1 lists the radioactive isotopes used for studies of pre-solar grains \citep{Groopman:2015}.
Studies of pre-solar dust grain compositions have lead to the distinctions of grain origins from AGB stars, from supernovae, and from novae, all of which are copious producers of dust particles. Formation of stardust occurs in circumstellar environments where temperatures are cool enough \citep[e.g.][for a recent review of the open issues]{Cherchneff:2017}. On their journey  through the interstellar medium, heating and partial or complete destruction may occur from starlight or even shocks from supernovae \citep{Zhukovska:2016}.  Also a variety  chemical and physical reactions may reprocess dust grains \citep{Dauphas:2016}. Thus, the journey from the stardust source up to inclusion in meteorites which found their way to Earth remains subject to theoretical modelling and much residual uncertainty \citep{Jones:2009}. 
Nevertheless, cosmic dust particles are independent astrophysical messengers, and complement studies based on electromagnetic radiation in important ways. 
Grain composition and morphology from the stardust laboratory measurements are combined with astronomical results such as characteristic spectral lines (e.g. from water ice, or a prominent feature associated with silicate dust), and interpreted through (uncertain) theories of cosmic dust formation and transport \citep{1998AREPS..26..147Z,Cherchneff:2016}.
Experimental difficulties and limitations arise from sample preparation through a variety of complex chemical methods, and by the extraction techniques evaporising material from the dust grain surfaces for subsequent mass spectrometry (see Chapter 10). 
\begin{figure}
  \centerline{
  \includegraphics[width=0.32\textwidth]{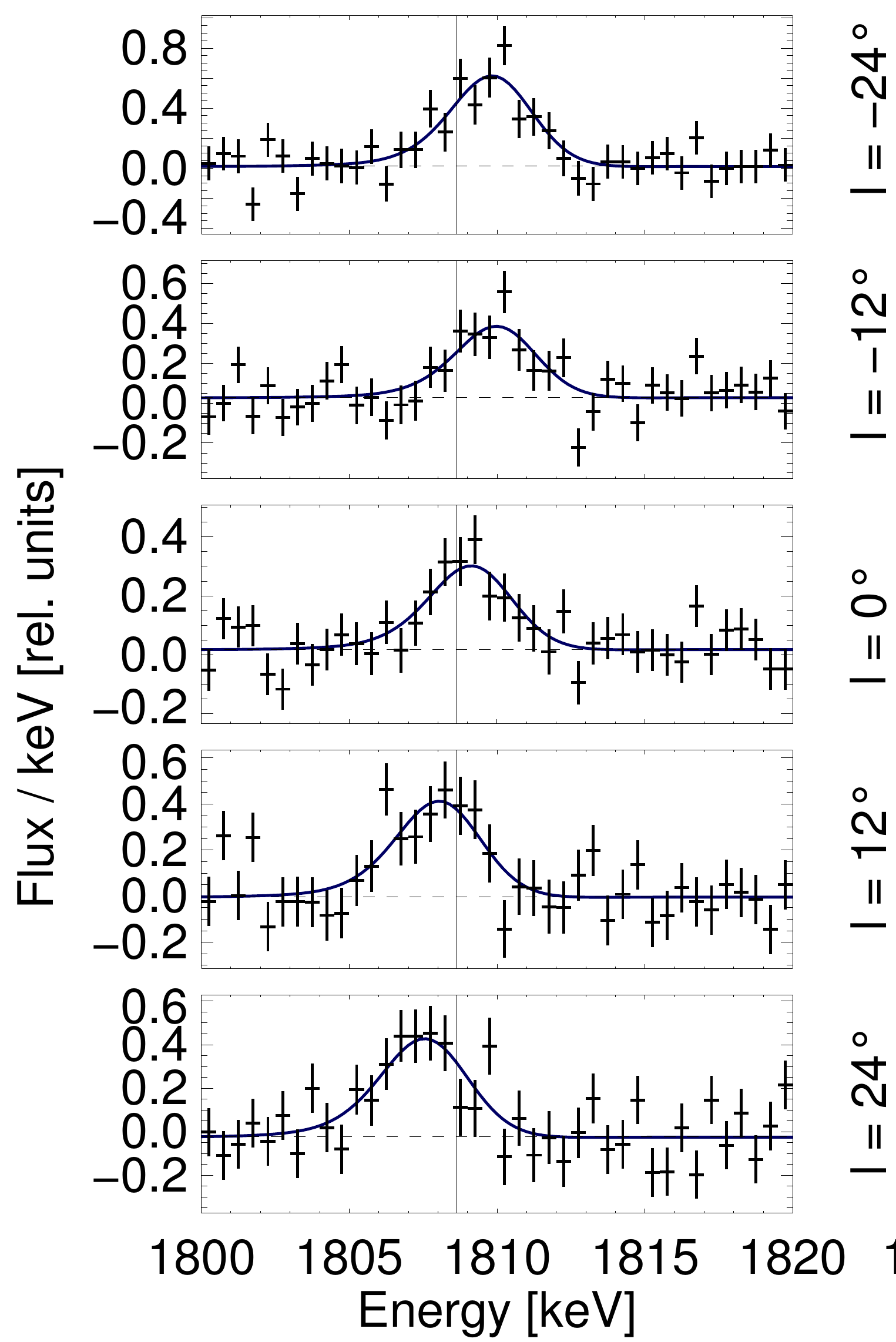} 
   \includegraphics[width=0.67\textwidth]{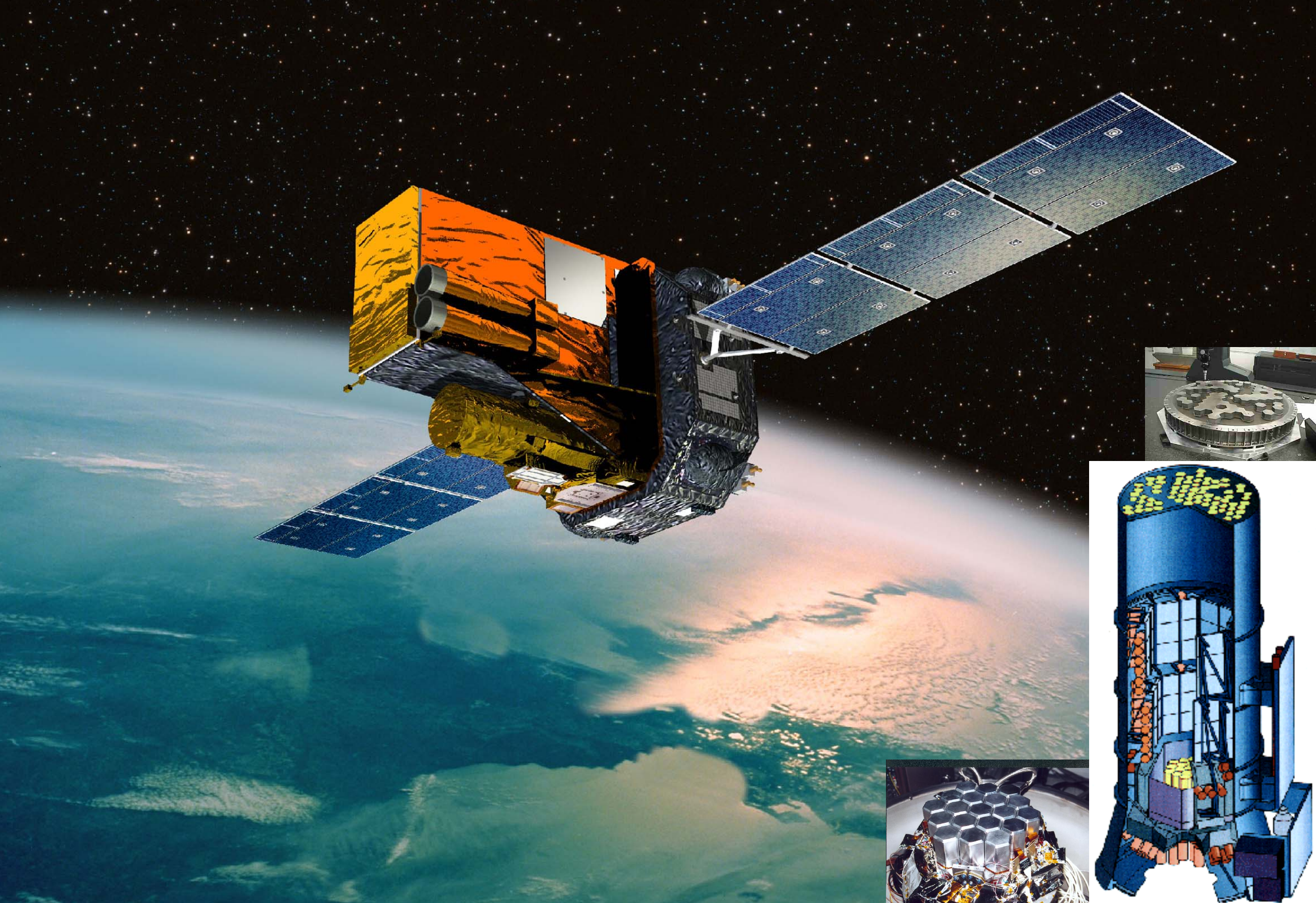}} 
  \caption{Example of a  $\gamma$-ray line measurement: The characteristic line from $^{26}$Al decay at 1808.63 keV appears Doppler-shifted from large scale galactic rotation, as it is viewed towards different galactic longitudes \citep[left; from][]{Kretschmer:2013}. 
 This measurement was performed with the SPI spectrometer on INTEGRAL, an example of a present-generation space-borne $\gamma$-ray telescope. The INTEGRAL satellite \index{INTEGRAL} (artist view picture, ESA) has two main telescopes;  the spectrometer SPI, one of them, is shown at the lower-right schematically with its 19-detector Ge camera and the tungsten mask for imaging by casting a shadow onto the camera.  Space-based instruments of this kind are required to directly measure the characteristic $\gamma$-ray lines from the decay of unstable isotopes near sites of current-epoch cosmic element formation.}
 \label{fig_1_gamma}
\end{figure}
\begin{table*}
\begin{tabular}{|l|c|c|c|c|}
  \hline 
  \quad\quad {\bf Decay} & {\bf Lifetime} & {\bf$\gamma$-ray Energy} & {\bf Site} & {\bf Process } \\
   \quad\quad {\bf Chain} & [y] & [keV] &         & {\bf Type } \\
                     &     & (branching ratio [\%]) & (detections) &   \\
    \hline \hline
    \, $^{7}$Be\,$\rightarrow ^{7}$Li    \,       &   0.21         &      478  (100)  &     Novae           & explosive  \\
                                                                            &                    &                           &                           & H burning  \\
   \hline 
    \, $^{56}$Ni$\longrightarrow ^{56}$Co$\longrightarrow^{56}$Fe\,   &   0.31         &   847 (100), 1238 (68)             &     SNe           & NSE  \\
                                                                            &                    &                 2598 (17), 1771 (15) &  (SN1987A,            & burning  \\
                                                                            &                    &                 &  (SN1991T, SN2014J)         &   \\  
                                                                                                                                                      &                    &                 {and 511 from e$^+$} &           &   \\
   \hline 
   \,  $^{57}$Co$\longrightarrow ^{57}$Fe       \,      &   1.1         &   122 (86), 136 (11)             &     SNe           & NSE  \\
                                                                            &                    &                 &  (SN1987A         & burning  \\
   \hline 
   \,  $^{22}$Na$\longrightarrow ^{22}$Ne   &   3.8         &   1275 (100)             &     Novae           & explos.  \\
                                                                            &                    &             {and 511 from e$^+$}        &           & H burning  \\
  \hline 
   \,  $^{44}$Ti$\longrightarrow ^{44}$Sc$\longrightarrow^{44}$Ca\,   &   89         &   68 (95), 78 (96)             &     SNe           & NSE   \\
                                                                            &                    &                 1156 (100) &  (Cas A, SN1987A)           & $\alpha$ freeze-  \\
                                                                            &                    &             {and 511 from e$^+$}        &           & out  \\
   \hline 
\,     $^{26}$Al$\longrightarrow ^{26}$Mg   &   1.04 10$^6$         &   1809 (100)           &     ccSNe, WR           & H burning  \\
                                                                            &                    &                 &  Novae, AGB (Galaxy)           & ($\nu$-proc.)  \\
                                                                            &                    &           {and 511 from e$^+$}       &  (Cygnus;Sco-Cen;          &  \\
                                                                            &                    &              &  Orion; Vela)           &  \\
   \hline 
    \, $^{60}$Fe$\longrightarrow ^{60}$Co$\longrightarrow^{60}$Ni\,   &   3.8 10$^6$         &   1173 (100), 1332 (100)             &     SNe           & He,C  \\
                                                                            &                    &                 59 (2) &  (Galaxy)           & shell burning  \\
   \hline 
 \,    e$^{+}\longrightarrow $Ps,..$\longrightarrow\gamma\gamma(\gamma)$ \,  &   $\sim$10$^7$        &   2$\cdot$511 ($\sim$100), cont $<$510  &     radioactivities      & $\beta^+$ decay  \\
                                                                            &                    &                   &  Pulsars, $\mu$QSOs, ...           &  rel. plasma  \\
                                                                            &                    &                   &  (Galactic bulge; disk)           &    \\
  \hline 
\end{tabular}
\caption{Radioactivities with gamma-ray line emission, sorted by ascending radioactive mean lifetime \cite[updated from][]{Diehl:2006g}. }
\label{fig:TableRadioactGammaLines}
\end{table*}
\item
Characteristic $\gamma$-ray lines from cosmic sources were not known until the 1960$^{ies}$, when space flight and its investigations of the near-earth space radiation environment had stimulated measurements of $\gamma$-rays. The discovery of a cosmic $\gamma$-ray line feature near 0.5~MeV from the direction towards the center of our Galaxy in 1972 \citep{1972ApJ...172L...1J} stimulated balloon and satellite experiments for cosmic $\gamma$-ray line spectroscopy. 
Radioactive isotopes are ejected into the surroundings of their nucleosynthesis sources, and become observable through their gamma-ray line emission once having left dense production sites where not even gamma-rays may escape. Nuclear gamma-rays can penetrate material layers of integrated thickness of a few grams~cm$^{-2}$. A typical interstellar cloud would have $\sim$0.1~g~cm$^{-2}$, SNIa envelopes are transparent to gamma-rays after 30--100~days, depending on explosion dynamics.  Depending on radioactive lifetime, gamma-ray lines measure isotopes which originate from single sources (the short-lived isotopes) or up to thousands of sources as accumulated in interstellar space over the radioactive lifetime of long-lived isotopes (see Table~1.2). %

Decay of the isotopes $^{26}$Al, $^{60}$Fe, $^{44}$Ti, $^{57}$Ni, and $^{56}$Ni \index{isotopes!26Al} \index{isotopes!44Ti} \index{isotopes!60Fe} \index{isotopes!57Ni} \index{isotopes!56Ni} in distant cosmic sites is an established fact, and astrophysical studies make use of such measurements. The downsides of those experiments is the rather poor resolution by astronomy standards (on the order of degrees), and the sensitivity limitations due to large instrumental backgrounds, which effectively only shows the few brightest sources of cosmic $\gamma$-rays until now \citep[see][for a discussion of achievements and limitations]{Diehl:2006g}. 
\end{itemize}

\noindent
Despite their youth and limitations, both methods to address cosmic radioactivities share a rather direct access to isotopic information, unlike other fields of astronomy. 
Isotopic abundance studies in the nuclear energy window will be complemented for specific targets and isotopes from the new opportunities in optical and radio/sub-mm spectroscopy. 
From a combination of all available astronomical methods, the study of cosmic nucleosynthesis will continue to advance towards a truly astrophysical decomposition of the processes and their interplays. This book describes where and how specific astronomical messages from cosmic radioactivity help to complement these studies.


\bibliographystyle{aa}

\begin{figure}
  \includegraphics[width=0.35\textwidth]{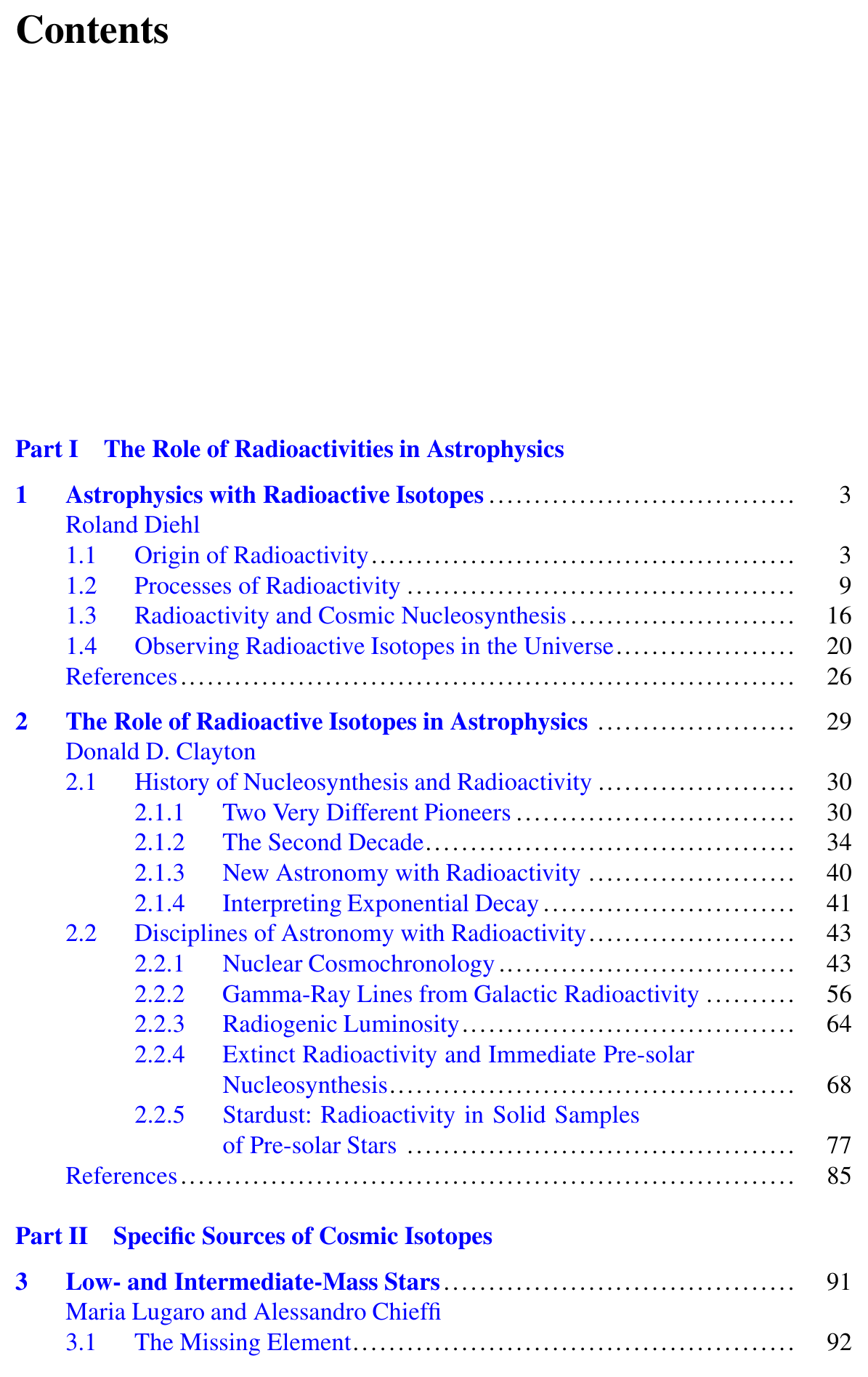}  \includegraphics[width=0.35\textwidth]{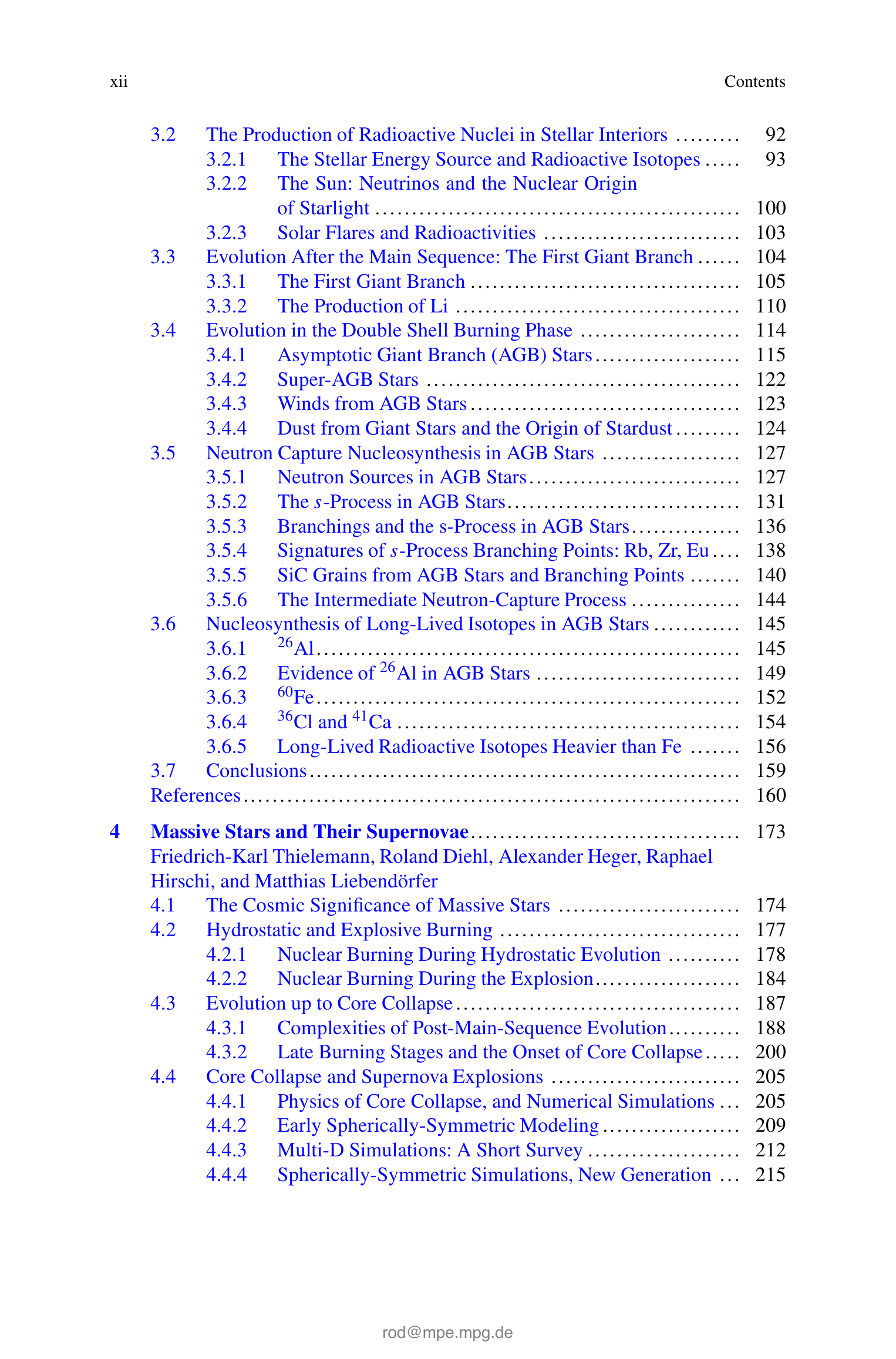}
  \includegraphics[width=0.35\textwidth]{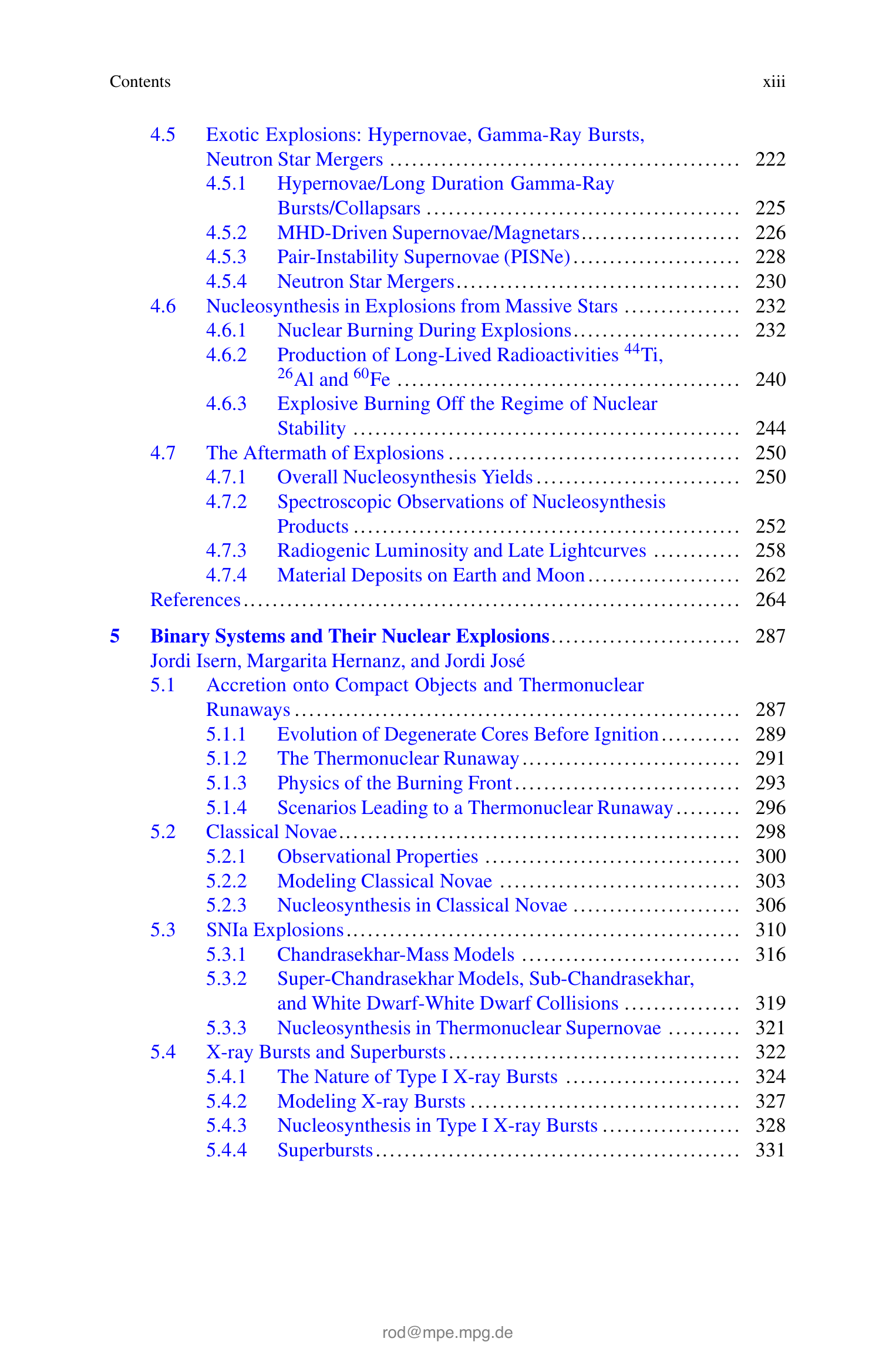}  \includegraphics[width=0.35\textwidth]{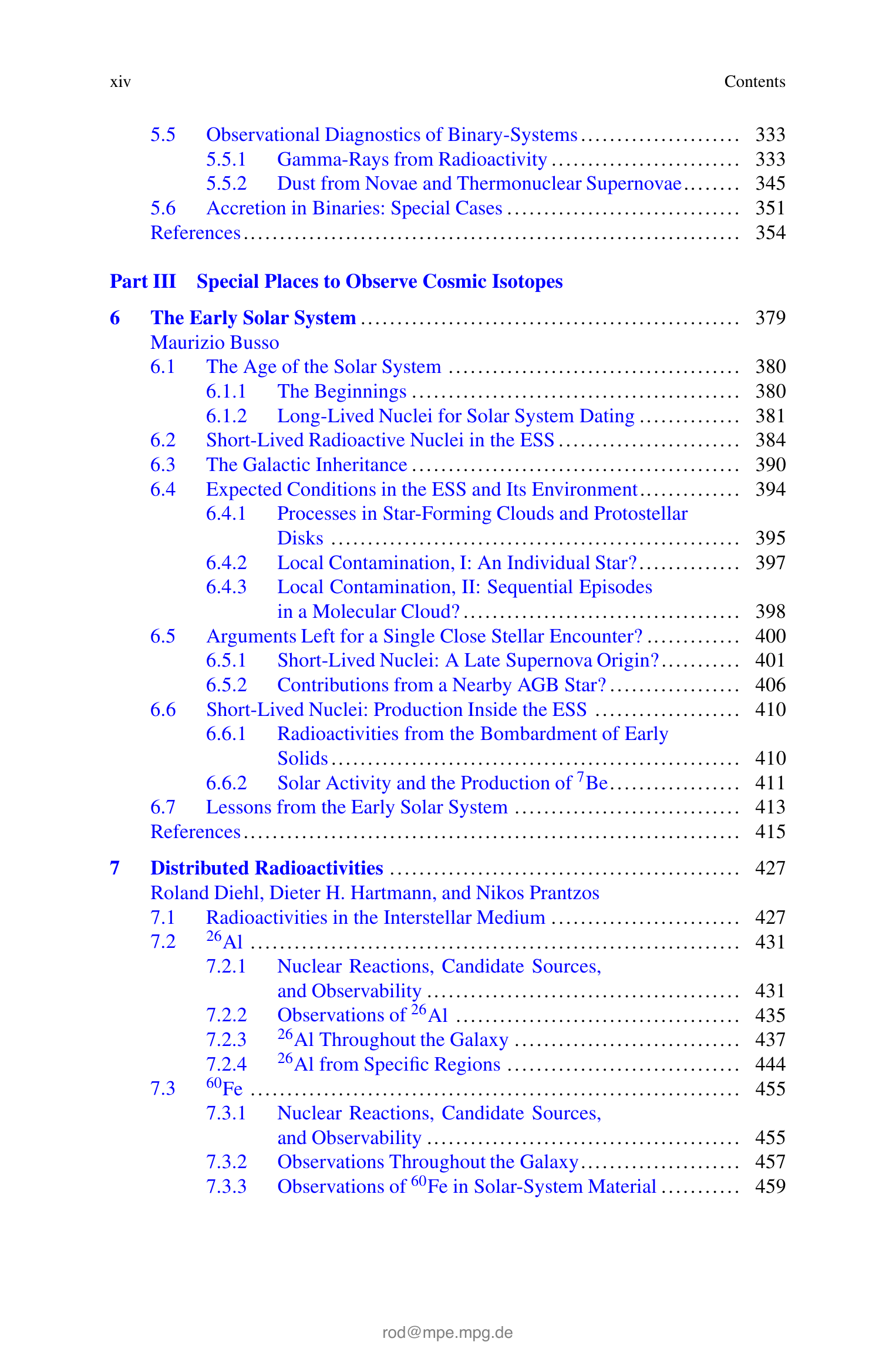}
  \includegraphics[width=0.35\textwidth]{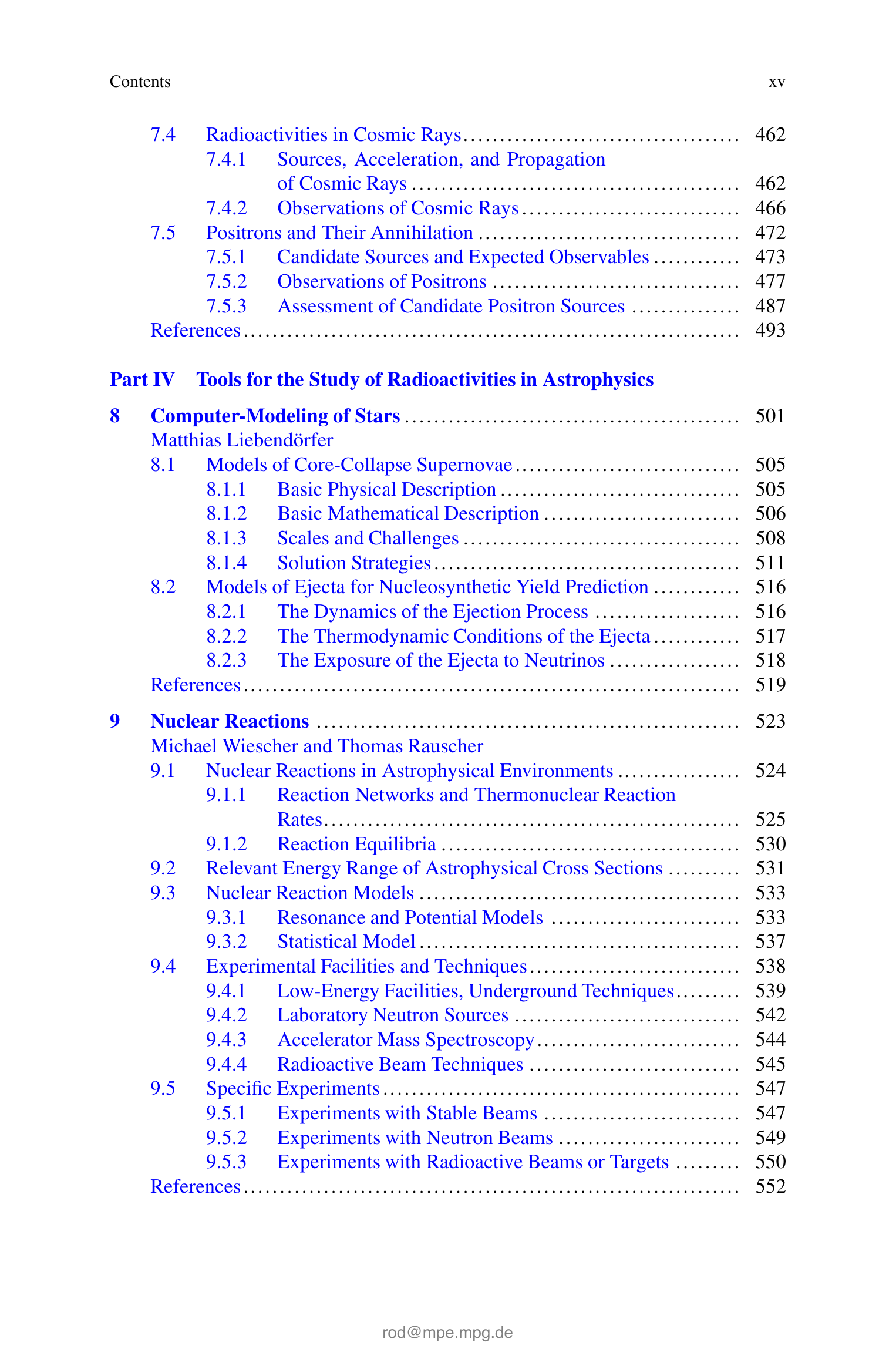}  \includegraphics[width=0.35\textwidth]{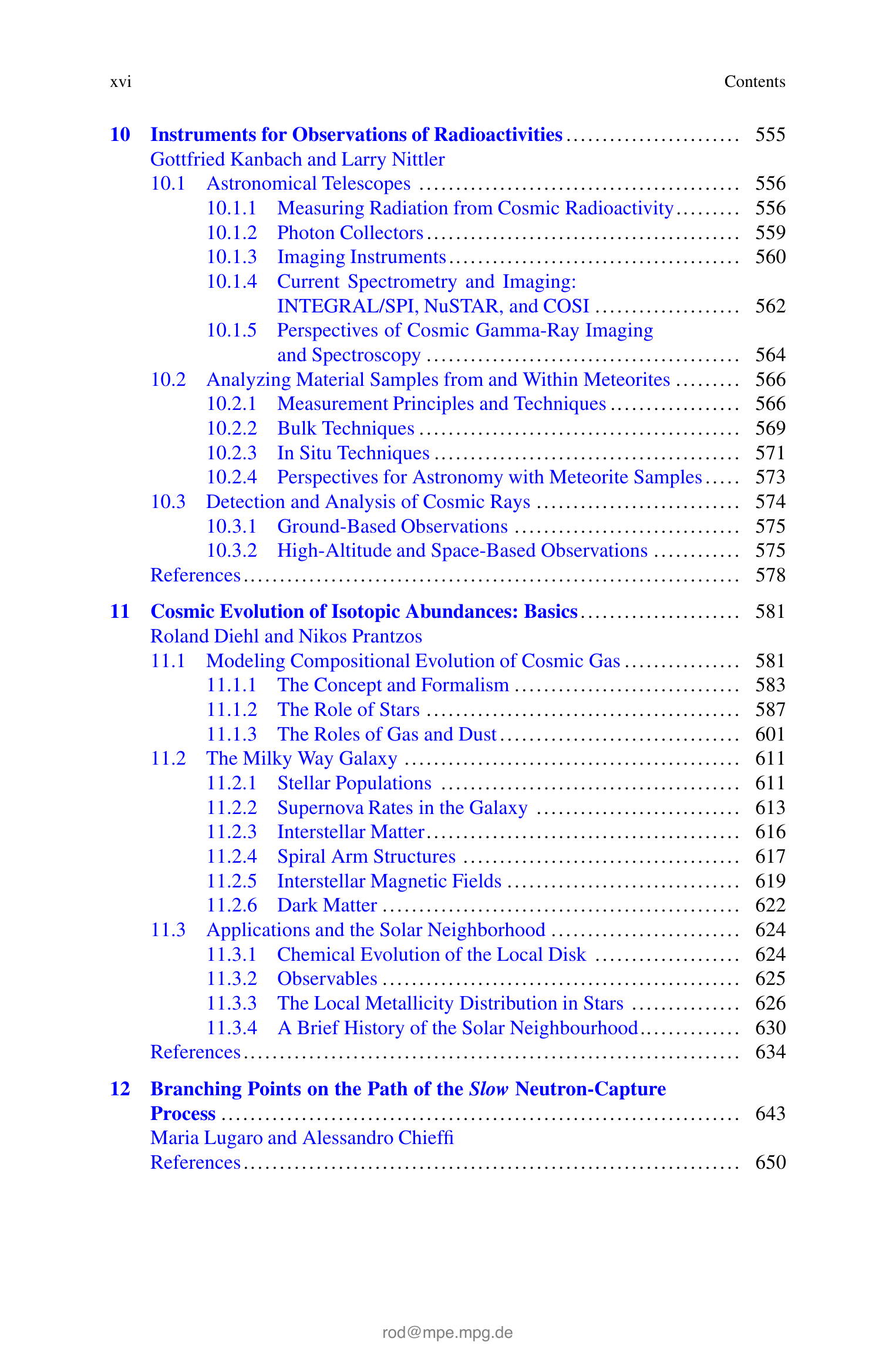}
  \includegraphics[width=0.35\textwidth]{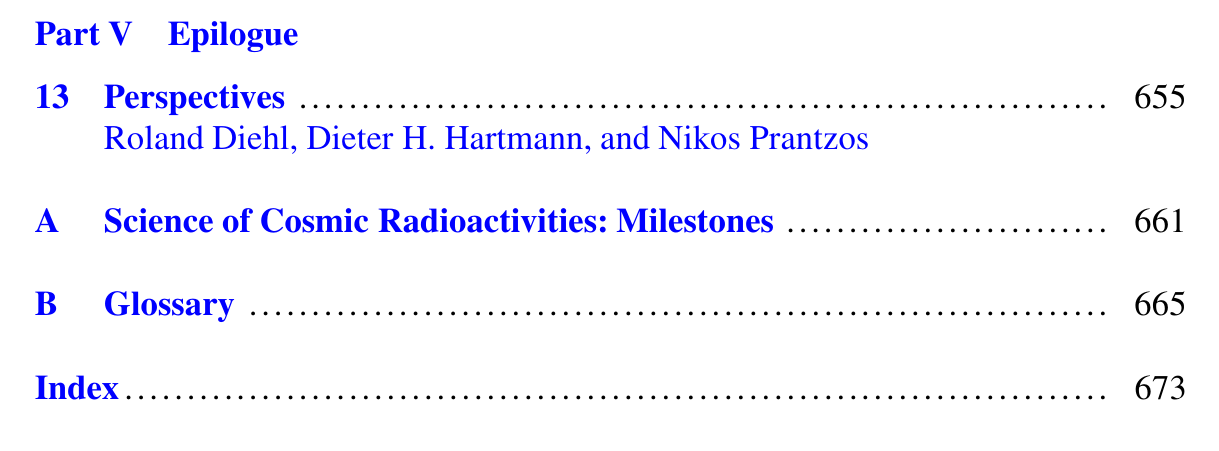} 
  \caption{Table of contents: "Astrophysics with radioactive isotopes", Springer ASSL 453 (2018) }
 \label{fig_TOC}
\end{figure}
\end{document}